\documentstyle[12pt]{article}

\def\ie{\hbox{\it i.e.}}

\def\Ups{\Upsilon}

\def\fun{\mbox{Fun$(G_{q})$}}
\newcommand{\tr}{\triangleright}
\def\cross{\mbox{$\times \!\rule{0.3pt}{1.1ex}\,$}}
\def\smash{{\A \cross \U}}
\def\R{\mbox{$\cal R$}}
\def\A{\mbox{$\cal A$}}
\def\U{\mbox{$\cal U$}}
\newcommand{\DA}{\Delta _{\cal A}}
\newcommand{\AD}{{}_{\cal A}\Delta }
\newcommand{\UD}{{}_{\cal U}\Delta }

\def\z{\hspace*{9mm}}
\def\x{\hspace{3mm}}
\newcommand{\ad}{\stackrel{\mbox{\scriptsize ad}}{\triangleright}}
\newcommand{\da}{\stackrel{\mbox{\scriptsize ad}}{\triangleleft}}
\def\I{\mbox{\boldmath $i$}}
\def\Ix#1{\mbox{\boldmath $i$}_{\chi_#1}}
\def\Li{\hbox{\large\it \pounds}}
\def\Lix#1{\hbox{\large\it \pounds}_{\chi_#1}}
\def\Lio#1#2{\hbox{\large\it \pounds}_{O_#1{}^#2}}
\def\dl{\mbox{\bf d}}
\newcommand{\om}{\mbox{$\omega $}}
\newcommand{\al}{\mbox{$\alpha $}}
\newcommand{\fum}{\mbox{Fun({\bf M}$_{q}$)}}
\newcommand{\tqm}{\mbox{${\cal T}(\mbox{\bf M}_q)$}}

\def\tq{\mbox{${\cal T}_q$}}
\def\dg{\mbox{\boldmath$\delta $}}

\begin{document}
\begin{titlepage}
\begin{center}
December 8, 1993    \hfill   hep-th/9312076\\
          \vskip .25in
{\large \bf From Quantum Planes to Quantum Groups and back;
Cartan Calculus}\footnote{This
work was supported in part by the Director, Office of Energy Research,
Office of High Energy and Nuclear Physics, Division of High Energy
Physics of the U.S. Department of Energy under Contract
DE-AC03-76SF00098 and in part by the National Science Foundation under
grant PHY-90-21139.}
\vskip .25in
Peter Schupp\footnote{schupp@physics.berkeley.edu}
\vskip .25in
{\em Department of Physics\\
University of California\\
and\\
Theoretical Physics Group\\
Lawrence Berkeley Laboratory\\
University of California\\
Berkeley, California 94720}
\vskip .25in
{\small\bf Abstract\vspace{-5mm}}
\end{center} {\small
A Cartan Calculus of Lie derivatives, differential forms, and inner
derivations, based on an undeformed Cartan identity, is constructed.
We attempt a classification of various types of quantum Lie algebras and
present a fairly general example for their construction, utilizing pure braid
methods, proving orthogonality of the adjoint representation and giving
a (Killing) metric and the quadratic casimir. A reformulation of the Cartan
calculus as a braided algebra and its extension to quantum planes,
directly and induced from the group calculus, are provided.}
\end{titlepage}
\newpage
\renewcommand{\thepage}{\arabic{page}}
\setcounter{page}{1}
\tableofcontents \newpage

\section{Introduction}

A classical plane can be fully described by the commutative algebra of
(coordinate) functions over it. This algebra is typically covariant
under the action of some symmetry group, and derivatives on it satisfy
an undeformed product rule.
A quantum plane in contrast to this is
covariant under a quantum group whose non-commutative algebra of
functions \A\  also implies a non-commutative
algebra of functions on the q-plane \fum.
The transformations of \fum\
and of the dual algebra of quantum derivatives $\tqm$ is most
easily described in terms of \A-coactions on coordinate functions
and partial derivatives:
\begin{eqnarray}
\DA x^{i} & = & x^{j} \otimes S t^{i}{}_{j},\\
\DA \partial _{i} & = & \partial _{j} \otimes S^{2} t^{j}{}_{i},
\end{eqnarray}
which we sometimes write in short matrix form as
$x  \to   t^{-1}\cdot  x$ and
$\partial   \to   \partial \cdot S^{2}t$.
The antipode ``$S$'' was inserted here to make these transformations
{\em right} coactions, the $S^{2}$ is needed for covariance (see below).\\
{\em Remark:} One can use $t^j{}_i$ in place of $S t^i{}_j$. Then $x \to x
\cdot t$ and
$\partial \to St \cdot \partial$. The choice is purely conventional.

\subsection{Product Rule for Quantum Planes}

Having made the ring of functions non-commutative,
we must now also modify the product rule in order to
retain covariant equations.
We make the following ansatz (see \cite{WZ})
\begin{equation}
\partial _{i} x^{k} = \partial _{i}(x^{k}) + L_{i}{}^{j}(x^{k}) \partial _{j},
\label{prodrule}
\end{equation}
where $\partial _{i}(x^{k}) = \delta^k_i$ and
$L_{i}{}^{j}$ is a linear operator that describes the braiding of $\partial
_{i}$
as it moves through $x^{k}$.
In place of the coordinate function
$x^{k}$ one could write any other
function in \fum\ and in particular (formal) power series in the
coordinate functions. When we consider products of coordinate
functions we immediately see that $L$ satisfies
\begin{equation}
L_{i}{}^{j}(x y) = L_{i}{}^{l}(x) L_{l}{}^{j}(y),\z L_{i}{}^{j}(1)
= \delta _{i}^{j},
\end{equation}
which can be reinterpreted in Hopf algebra language as
$\Delta  L = L \dot{\otimes} L$ and $\epsilon (L) = I$; $S L = L^{-1}$ follows
naturally.
We are hence let to believe  that  $L$ should belong to  some Hopf algebra,
the Braiding Hopf Algebra.
In the case of linear quantum groups $L$ is for instance
an element of the quasitriangular
Hopf algebra $\U$ of the quantum symmetry group.
Considering multiple derivatives gives additional conditions that
can be summarized by requiring that
\begin{equation}
\UD \partial _{i} = L_{i}{}^{j} \otimes \partial _{j}
\end{equation}
be a Hopf algebra coaction, {\em i.e.}
\begin{equation}
\UD(\partial  \partial ') = \UD(\partial ) \UD(\partial '),\x
(i\!d \otimes \UD) \UD
= (\Delta  \otimes i\!d) \UD,\x (\epsilon  \otimes i\!d) \UD = i\!d.
\end{equation}
For arbitrary functions $f$ and derivatives $\partial $ we
find a generalized product rule
\begin{equation}
\fbox{$\partial  f = \partial (f) + \partial _{1'}(f) \partial _{2}$},
\end{equation}
where ${}_{\cal U}\Delta  \partial
\equiv  \partial _{1'} \otimes \partial _{2}$. Covariance of the product rule
(\ref{prodrule}) under coactions is expected to give
strong conditions on $L_{i}{}^{j}$.

\subsubsection[Covariance]{Covariance of: $\partial _{i} f
= \partial _{i}(f) + L_{i}{}^{j}(f) \partial _{j}$}
\label{covdfl}

We need to use an {\bf inductive} approach: We start by requiring that
\begin{equation}
\DA(\partial _{i}(x^{j})) = \DA \partial _{i}(\DA x^{j}).\z\mbox{\it (anchor)}
\end{equation}
This is in fact satisfied, because we
already have $\DA x^{j} = x^{l} \otimes S t^{j}{}_{l}$ and
iff $\DA \partial _{j} = \partial _{l} \otimes
S^{2} t^{l}{}_{j}$ then:
$\DA \partial _{i}(\DA x^{j})
= \partial _{k}(x^{l}) \otimes S^{2} t^{k}{}_{i} S t^{j}{}_{l}
= \delta _{k}^{l} \otimes S^{2} t^{k}{}_{i}
S t^{j}{}_{l} = \delta _{i}^{j} \otimes 1$, in agreement
with $\DA(\partial _{i}(x^{j})) = \DA(\delta _{i}^{j}) =
\delta _{i}^{j} \otimes 1$. That was the anchor;
now the induction to higher powers
in the coordinate functions:
Assume that the action of $\partial_i$ on $f$ is covariant:
\begin{equation}
\DA(\partial _{i}(f)) = \DA \partial _{i}(\DA f),\label{dapif}
\end{equation}
where $f$ is a function of the coordinate functions $x^{i}$.
Try to proof covariance of the $\partial_i$--$f$ commutation relation, \ie\
\begin{equation}
(\protect\ref{dapif})\:\stackrel{?}{\Rightarrow}\:\DA(\partial _{i} f)
= \DA(\partial _{i})\cdot \DA(f).\z\mbox{\it
(induction)}
\end{equation}
After some computation we find
\begin{equation}
\DA(L_i{}^j(f))(1 \otimes S^2t^k{}_j) \stackrel{!}{=}
L_l{}^k(f^{(1)}) \otimes S^2t^l{}_i f^{(2)'},\label{first}
\end{equation}
where $\DA(f) \equiv f^{(1)} \otimes f^{(2)'}$. This simplifies further if we
know how
$L_i{}^j$ acts on $f$. If the braiding Hopf algebra acts like the covariance
quantum group, then
$L_i{}^j(f) = f^{(1)}<L_i{}^j,f^{(2)'}>$, $L_i{}^j \in \A^*$ and
(\ref{first}) becomes
\begin{equation}
\left( L_{i}{}^{j}(f^{(2)'}) S^{2}t^{k}{}_{j}
- S^{2}t^{l}{}_{i} \widehat{L_{l}{}^{k}}(f^{(2)'}) \right)
\otimes f^{(1)} = 0,
\end{equation}
where $\widehat{\x}:\smash \to \smash$ is the projector onto right-invariant
vector fields \cite{thesis}: $\widehat{x} = S^{-1}(x^{(2)'}) x^{(1)}$
with $\DA(x) \equiv x^{(1)} \otimes x^{(2)}$, such that
$\widehat{L_l{}^k}(f^{(2)'}) =$ $<L_l{}^k,f^{(2)'}> f^{(3)'}$.
This is satisfied if
\begin{equation}
L_{i}{}^{j}(a) S^{2}t^{k}{}_{j}
= S^{2}t^{l}{}_{i} \widehat{L_{l}{}^{k}}(a),\z\forall a \in \A.
\end{equation}
(The reverse is true only if \A\ is generated by  $[S t^{i}{}_{j}]$ --- or
$[t^i{}_j]$, if the choose the
convention $\DA(x) = x \cdot t$.)
In the case where the braiding Hopf algebra is quasitriangular, there
are (exactly) two natural choices
\begin{equation}
L_{j}{}^{i} \propto \left\{\begin{array}{l} S^{-1} L^{-}{}^{i}{}_{j}
\equiv  <\R,S^{2}t^{i}{}_{j} \otimes i\!d>\\
S^{-1} L^{+}{}^{i}{}_{j} \equiv  <\R,i\!d \otimes S
t^{i}{}_{j}>\end{array}\right.
\end{equation}
that satisfy the above equation and all other requirements
(coproduct, {\em e.t.c.}).

For the Wess-Zumino quantum plane \cite{WZ} the action of $L$ on the
coordinate functions is linear and of first degree in those functions,
so we can use the coaction $\DA$ to express it:
\begin{equation}
L_{i}{}^{j}(x^{k}) = <L_{i}{}^{j},S t^{k}{}_{l}> x^{l} \propto
\left\{\begin{array}{l} r^{kj}{}_{li} x^{l}\\
(r^{-1})^{jk}{}_{il} x^{l}\end{array}\right.\label{lplane}
\end{equation}
in perfect agreement with \cite{WZ}.
(The overall multiplicative constant ($\frac{1}{q}$) is not fixed by covariance
considerations but is given by the characteristic equation of $\hat{r}$
and the requirement that $\widehat{C}^{kj}{}_{li} \equiv <L_i{}^k,St^j{}_l>$
should have an eigenvalue $-1$.)

\subsection{Quantum Groups}

A quantum group is a quantum plane covariant under itself. However, it has more
structure and the coactions $\DA$ and $\UD$ are now completely
determined by the multiplication in \U\ and \A: Let $\phi  \in \A\cross\U$
and $\DA \phi  \equiv  \phi ^{(1)} \otimes \phi ^{(2)'}$; then
\begin{equation}
        \Lix{{}}(\phi ) = \chi  \ad \phi  \equiv  \chi _{(1)} \phi  S \chi
_{(2) }
        = \phi ^{(1)} <\chi ,\phi ^{(2)'}>,\z\forall \chi  \in \U
\end{equation}
determines $\DA$. The coaction $\UD$ is simply the coproduct
$\Delta :\U \to  \U \otimes \U$, so that the product rule becomes
\begin{equation}
x a = a_{(1)} <x_{(1)},a_{(2)}> x_{(2)},
\end{equation}
where $x \in \U,\: a \in \A$. This defines the multiplicative structure
in the so called cross product algebra \cite{SWZ3} $\smash$.
Interestingly, equation (\ref{lplane}) does not apply in the
case of a quantum
group: In that case $t$ is replaced by the
adjoint representation $T$ and $L$ becomes $O$, a part in the coproduct
of the basic generators.
Not all elements of $T$ are linearly
independent. There is a trivial partial sum $T^{(ii)}{}_{(kl)} = 1 \delta
_{(kl)}$;
the same sum for $O$, $O^{(ii)}{}_{(kl)} =: Y_{(kl)}$, is in general
non-trivial
thus leading to a contradiction. An explanation for
this is that quantum groups have more structure than quantum planes.
They already contain an intrinsic braiding and do not leave any freedom
for external input such as $\R$ in equation (\ref{lplane}); the product rule is
in fact automatically
covariant by the construction of the cross product algebra. There are, however,
some indications that $O$ and $T$ might be related to a universal
$\tilde{\R}$ that lives in the sub-Hopf algebra of \A\, generated by the
elements of $T$.

{}From the discussion of the quantum planes we would like to keep the idea
of a finite number of so-called bicovariant generators $\chi _{i}$ that
close under adjoint action
        $\chi _{i} \ad \chi _{j} = \chi _{k} f_{i}{}^{k}{}_{j}$
and span an invariant subspace of \U, $i.e.$
        $\DA \chi _{j} = \chi _{k} \otimes T^{k}{}_{j}$.
We  call quantum groups with such generators Quantum Lie
Algebras.
In following section we will give  more
precise definitions of quantum Lie algebras.

\section{Cartan Calculus}

\subsection{Cartan Identity}

The central idea behind Connes Universal Calculus \cite{Co1} in the context of
non-com\-mu\-ta\-tive geometry was to retain from the classical
differential geometry
the nilpotency of $\dl$
\begin{equation}
\dl^{2} = 0
\end{equation}
and the undeformed Leibniz rule
for $\dl$\footnote{We use parentheses
to delimit operations like \dl, $\I_{x}$ and $\Li_{x}$, {\em e.g.}
$\dl a = \dl(a)
+ a \dl$.  However, if the limit of the operation is clear from the context,
we will
suppress the parentheses, {\em e.g.} $\dl(\I_{x} \dl a) \equiv
\dl(\I_{x}(\dl(a)))$.}
\begin{equation}
\dl \alpha  = \dl(\alpha ) + (-1)^{p} \alpha  \dl    \label{LEIBNIZ}
\end{equation}
for any $p$-form $\alpha $. The exterior derivative $\dl$ is a scalar making
this equation hard to deform, except for a possible multiplicative
constant in the second term.
Here we want to base the construction of a differential calculus on
quantum groups on two additional classical formulas: To extend the definition
of a Lie derivative from functions and vector fields to forms
we postulate
\begin{equation}
\Li \circ \dl = \dl \circ \Li; \label{LIE}
\end{equation}
this is essential for a geometrical interpretation.
The second formula that we can --- somewhat surprisingly ---
keep undeformed in the quantum case is 
\begin{equation}
\Lix{i} = \Ix{i} \dl + \dl \Ix{i},\z\mbox{\it (Cartan
Identity)}\label{CARTAN}
\end{equation}
where $\chi _{i}$ are the generators of some quantum Lie algebra.
The only possibility to deform this equation and not violate its
covariance is to introduce multiplicative deformation parameters
$\kappa ,\lambda $ for the two terms on the right hand side of (\ref{CARTAN})
such
that now $\Lix{i} = \kappa  \Ix{i} \dl + \lambda  \dl \Ix{i}$. For a function
$a \in \A$ that gives
$$\Lix{i}(a) = \kappa  \Ix{i}(\dl a)$$
($\Ix{i}$ vanishes
on functions), for $\dl a$ we find
$$\Lix{i}(\dl a) = \lambda  \dl(\Ix{i}(\dl a))$$
and finally together
$$\Lix{i}(\dl a) = \frac {\lambda }{\kappa } \dl(\Lix{i}(a)),$$
in contrast to (\ref{LIE}) unless $\frac {\lambda }{\kappa } = 1$, in which
case we can easily
absorb either $\kappa $ or $\lambda $ into $\Ix{{}}$. Being now (hopefully)
convinced
of our two basic equations (\ref{LIE}) and (\ref{CARTAN}) we want
to turn to the generators $\chi _{i}$ next.

\section{Quantum Lie Algebras}

A quantum Lie algebra is a Hopf algebra \U\ with a finite-dimensional
biinvariant sub vector space $\tq$ spanned by generators $\{\chi _{i}\}$
with coproduct
\begin{equation}
\Delta  \chi _{i} = \chi _{i} \otimes 1 + O_{i}{}^{j} \otimes \chi _{j}.
\end{equation}
More precisely we will call this a quantum Lie algebra of {\bf type II}.
Let $\{\omega ^{j} \in \tq^{*}\}$ be a dual basis of 1-forms corresponding to a
set of
functions $b^{j} \in \A$ via $\omega ^{j} \equiv  S b^{j}_{(1)} \dl
b^{j}_{(2)}$; {\it i.e.}
\begin{eqnarray}
\AD(\chi _{i}) &=& 1\otimes \chi _{i}, \nonumber \\
\DA(\chi _{i}) &=& \chi _{j} \otimes T^{j}{}_{i},\x T^{j}{}_{i}\in\fun,\\
\I_{\chi _{i}}(\omega ^{j}) &  = & -<\chi _{i},S b^{j}> = \delta
^{j}_{i},\label{ebdual}\\
\AD(\omega ^{i}) &=& 1\otimes \omega ^{i},\\
\DA(\omega ^{i}) &=& \omega ^{j} \otimes S^{-1} T^{i}{}_{j}.
\end{eqnarray}
If the functions $b^{i}$ also close under adjoint coaction $\Delta ^{Ad}(b^{i})
= b^{j}
\otimes S^{-1} T^{i}{}_{j}$, we will call the corresponding quantum Lie algebra
one of {\bf type I}.
Getting a little ahead of ourselves let us mention that we can derive
an expression for the
exterior derivative of a function from the Cartan identity
(\ref{CARTAN}) in terms of these bases
\begin{equation}
\dl(a) = \omega ^{i}(\chi _{i} \tr a) = \omega ^{i} \Lix{i}(a) \label{exder}
\end{equation}
and that this leads to the following $f-\omega $ commutation relations
\cite{W2}
\begin{equation}
f \omega ^{i} = \omega ^{j} (O_{j}{}^{i} \tr f).
\end{equation}

\subsection{Generators, Metrics and the Pure Braid Group}

How does one practically go about finding the basis of generators $\{\chi
_{i}\}$
and the set of functions $\{b^{i}\}$ that define the basis of
1-forms $\{\omega ^{i}\}$?
Here we would like to present a method that utilizes pure braid group
elements as introduced in \cite{SWZ3}.

\subsubsection{Invariant Maps and the Pure Braid Group}

A pure braid element $\Ups$ is an element of
the pure braid group $B_2 \subset \U\hat{\otimes}\U$
that --- by definition --- commutes with all coproducts of
elements of $\U$, $i.e.$
\begin{equation}
\Ups \Delta (y) = \Delta (y) \Ups,\z \forall y \in \U.
\end{equation}
$\Ups$ maps elements of \A\ to left-invariant elements of \U\ with special
transformation properties under the right coaction:
\begin{equation}
\begin{array}{c}
\Ups:\A \to  \U\,:\x b \mapsto \Ups_{b} \equiv <\Ups,b \otimes
i\!d>,\\
\DA(\Ups_{b}) = \Ups_{b_{(2)}} \otimes S(b_{(1)}) b_{(3)} = <\Ups \otimes i\!d,
\tau ^{23}(\Delta ^{Ad}(b) \otimes i\!d)>,
\end{array}
\end{equation}
as we will now show.

\noindent {\bf Lemma: } {\em Let $\Ups \equiv \Ups_{i} \otimes \Ups^{i}
\in \U \otimes \U$ such that
$\Ups \Delta (x) = \Delta (x) \Ups$ for all $x \in \U$, then it follows that
$\: \Ups_{i} \otimes (x \ad \Ups^{i}) = (\Ups_{i} \da x) \otimes \Ups^{i}$
with $\Ups_{i} \da x \equiv S(x_{(1)}) \Ups_{i} x_{(2)}$ for all $x \in \U$.}

\noindent {\bf Proof:}
\begin{equation}
\begin{array}{rcl}
\Ups_{i} \otimes (x \ad \Ups^{i}) & \equiv &
             \Ups_{i} \otimes x_{(1)} \Ups^{i} S(x_{(2)})\\
& = & S(x_{(1)}) x_{(2)} \Ups_{i} \otimes x_{(3)} \Ups^{i} S(x_{(4)})\\
& = & S(x_{(1)}) \Ups_{i} x_{(2)} \otimes \Ups^{i} x_{(3)} S(x_{(4)})\\
& = & (\Ups_{i} \da x) \otimes \Ups^{i}.\z\Box
\end{array}
\end{equation}
For any function $b \in \A$, define
\begin{equation}
\Ups_{b} := <{\Ups},{b \otimes i\!d}> \: \in \U .
\end{equation}
{\bf Proposition:}
{\em This vector field has the following transformation property:}
\begin{equation}
\DA(\Ups_{b}) = \Ups_{b(2)} \otimes S(b_{(1)}) b_{(3)}.\label{proposition}
\end{equation}
{\bf Proof:}
\begin{equation}
\begin{array}{rcl}
\DA(\Ups_{b}) & = &<{\Ups_{i}},{b}> (e_{k} \triangleright \Ups^{i}) \
otimes f^{k}\\
&=& <{\Ups_{i} \triangleleft e_{k}},{b}> \Ups^{i} \otimes f^{k}\\
&=& <{\Ups_{i} \otimes e_{k}},{b_{(2)}>
    \otimes S(b_{(1)}) b_{(3)}} \Ups^{i} \otimes f^{k}\\
& = & \Ups_{b(2)} \otimes S(b_{(1)}) b_{(3)}.\z\Box
\end{array}
\end{equation}
The reverse statement is also true:\\
{\bf Proposition:} {\em If there is a linear map $\Ups:\A \rightarrow \U$,
realized and labeled by some element $\Ups \in \U \hat{\otimes} \U$ via
$b \mapsto \Ups_{b} \equiv <\Ups,b \otimes i\!d>,\x\forall b\in\A$, such that
the resulting element in \U\ transforms like
$\DA \Ups_{b} = \Ups_{b_{(2)}} \otimes S b_{(1)} b_{(3)}$;
then $\Ups \in B_{2}$,
{\em i.e.} $\Ups$ must commute with all coproducts.}\\
{\bf Proof:} For all $x \in \U$ and $b \in \A$
\begin{equation}
\begin{array}{rcl}
<\Delta x \Ups,b \otimes i\!d> & = & <\Delta x,b_{(1)}
             \otimes i\!d><\Ups,b_{(2)}\otimes i\!d>\\
        & = &<x_{(1)},b_{(1)}>x_{(2)}\Ups_{b_{(2)}}\\
        & = &<x_{(1)},b_{(1)}>\Ups_{b_{(3)}}<x_{(2)},S b_{(2)} b_{(4)}>
x_{(3)}\\
        & = &\Ups_{b_{(3)}}<x_{(1)},b_{(1)} S b_{(2)} b_{(4)}> x_{(2)}\\
        & = &\Ups_{b_{(1)}}<x_{(1)},b_{(2)}> x_{(2)}\\
        & = &<\Ups \Delta x,b\otimes i\!d>.\z\Box
\end{array}
\end{equation}
{}From this follows an important Corollary about invariant maps:\\
{\bf Corollary:} {\em If there exists a map $\phi :\A \rightarrow \A\cross\U$
such that $\DA \circ \phi  = (\phi  \otimes i\!d) \circ \Delta $; then it
follows that $\phi (b) = b_{(1)}<\Ups,b_{(2)}\otimes i\!d>$ with
$\Ups \in B_{2}$ for all
$b \in \A$ and vice versa.}\\
If we have a set of functions $\{b_i \in \A\}$ with
nice transformation properties, \ie\ closure under coadjoint action,
then we can use the maps described, to
construct vector fields corresponding
to --- and inheriting the simple transformation behavior of --- these
functions.
This construction can then be used to find a basis of vector fields that
closes under coaction and hence under (mutual) adjoint actions.

\subsubsection{Metric}

An element $\Ups$ of the pure braid group
defines a bilinear quadratic form on  the function Hopf algebra \A
\begin{equation}
(\:,\:): \A \otimes \A \to  k\,:\z a \otimes b \mapsto (a,b) = -<\Ups,a
\otimes S(b)> \in k,
\end{equation}
with respect to which we can construct orthonormal $(b_{i},b^{j}) = \delta
_{i}^{j}$
bases $\{b_{i}\}$ and $\{b^{j}\}$ of functions that in turn will define
generators $\chi _{i} := \Ups_{b_{i}}$  and 1-forms $\omega ^{j} :=
S(b^{j}_{(1)}) \dl b^{j}_{(2)}$.
Often one can choose span$\{b_{i}\}$ = span$\{b^{j}\}$; then one
starts by constructing one set, say $\{b_{i}\}$, of functions
that close under adjoint coaction
\begin{equation}
\Delta ^{Ad} b_{i} = b_{j} \otimes T^{j}{}_{i}.
\end{equation}
If the numerical matrix
\begin{equation}
\fbox{$\eta _{ij} := -<\Ups,b_{i} \otimes Sb_{j}>$}\z\mbox{\it (metric)}
\end{equation}
is invertible, $i.e.$ det$(\eta ) \neq 0$, then we can use its inverse
$\eta ^{ij} := (\eta ^{-1})_{ij}$ to raise indices
\begin{equation}
b^{i} = b_{j} \eta ^{ji}.
\end{equation}
This metric is invariant --- or $T$ is orthogonal --- in the sense
\begin{equation}
\begin{array}{rcl}
\eta _{ji}     & = & -<S\chi _{j},b_{i}>\\
        & = & -<S\chi _{j},b_{k}> ST^{k}{}_{l} T^{l}{}_{i}\\
        & = & -<\chi _{k},Sb_{l}> T^{k}{}_{j} T^{l}{}_{i}\\
        & = & \eta _{kl} T^{k}{}_{j} T^{l}{}_{i},
\end{array}
\end{equation}
where we have used the Hopf algebraic identity
\begin{equation}
<\DA(\chi ), Sa \otimes i\!d> = S(<S\chi  \otimes i\!d,\Delta ^{Ad}(a)>),
\end{equation}
which we will proof in the appendix.
Now we would like to show that we have actually
obtained a quantum Lie algebra of type I:\footnote{Note, that
$\Ups$ has to be carefully chosen to insure the correct number of
generators. Furthermore, we
still have to check the coproduct of the generators. If they are not of
the form $\Delta  \chi _{i} = \chi _{i} \otimes 1 + O_{i}{}^{j} \otimes \chi
_{j
}$ then we can still
consider a calculus with deformed Leibniz rule (see next section).}
\begin{equation}
-<\chi _{i},S b^{j}> =  -<\Ups,b_{i}
\otimes Sb_{k}> \eta ^{kj} = \eta _{ik}
\eta ^{kj} = \delta _{i}^{j},
\end{equation}
\begin{equation}
\DA(\chi _{i}) = \Ups_{b_{i(2)}} \otimes S(b_{i(1)}) b_{i(3)}
= \Ups_{b_{j}} \otimes T^{j}{}_{i} =
\chi _{j} \otimes T^{j}{}_{i}
\end{equation}
and
\begin{equation}
\Delta ^{Ad}(b^{i})
         =  b_{k} \otimes T^{k}{}_{j} \eta ^{ji}
         =  b_{k} \otimes \eta ^{kl} \eta _{ln} T^{n}{}_{j} \eta ^{ji}
         =  b^{k} \otimes S^{-1} T^{i}{}_{k}.
\end{equation}

\subsubsection{Casimir}

Once we have obtained a metric $\eta $, we can truncate the pure braid element
$\Ups$ and work instead with:
\begin{equation}
\Ups \to  \Ups_{trunc} =
-S(\chi _{i}) \otimes \chi ^{i} = -S(\chi _{i}) \otimes \chi _{j} \eta ^{ji},
\end{equation}
which also commutes with all coproducts.
One can construct casimir operators from elements of the pure
braid group. Let for instance $c:=\Ups^i S^{-1}(\Ups_i)$, where
$\Ups_i \otimes \Ups^i \equiv \Ups \in B_2$, then for any $y \in \U$:
\begin{equation}
\begin{array}{rcl}
S^{-1}(y) c = S^{-1}(y) \Ups^i S^{-1}(\Ups_i)
	& = & S^{-1}(y_{(3)}) \Ups^i y_{(2)} S^{-1}(y_{(1)}) S^{-1}(\Ups_i)\\
	& = & S^{-1}(y_{(3)}) \Ups^i y_{(2)} S^{-1}(\Ups_i y_{(1)})\\
	& = & S^{-1}(y_{(3)}) y_{(2)} \Ups^i S^{-1}(y_{(1)} \Ups_i)\\
	& = & \Ups^i S^{-1}(y \Ups_i)\\
	& = & \Ups^i S^{-1}(\Ups_i) S^{-1}(y) = c S^{-1}(y).
\end{array}
\end{equation}
For the truncated  pure braid element that gives the
quadratic casimir:
\begin{equation}
[\:\cdot  \circ \tau  \circ (S^{-1} \otimes i\!d)](\Ups_{trunc})
= \eta ^{ji} \chi _{j}
\chi _{i}.\z\mbox{\it (casimir)}
\end{equation}

\subsubsection{Example: The $r$-Matrix Approach}

Often one can take $b_{i} \in $span$\{t^{n}{}_{m}\}$,
where $t^{n}{}_{m}$ is a quantum
matrix in the defining representation of the quantum group under
consideration. If we are dealing with a quasitriangular Hopf algebra,
a natural choice for the pure braid element is
\begin{equation}
\Ups_{r} = \frac {1}{\lambda }\left(1 \otimes 1 - \R^{21} \R^{12}\right),
\end{equation}
where the term $\R^{21} \R^{12}$ has been introduced and extensively studied
by Reshetikhin \& Semenov-Tian-Shansky \cite{RSTS} and later by Jurco
\cite{J}, Majid \cite{pM}
and Schupp, Watts \& Zumino \cite{SWZ3}. These choices of $b_{i}$s and $\Ups$
lead to the $r$-matrix approach (see \cite{FRT})
to differential geometry on quantum
groups. The metric is
\begin{equation}
\eta  = -<X_{1},St_{2}>
= \frac {1}{\lambda }\left(\left[\left(r_{21}{}^{-1}\right)^{t_{2}}
\left(r_{12}{}^{t_{2}}\right)^{-1}\right]^{t_{2}} - 1 \right),
\label{eta}
\end{equation}
where $X_{1} = <\Ups_{r},t_{1} \otimes i\!d>$ and $r_{12}
= <\R,t_{1} \otimes t_{2}>$ not to be confused with the ``classical
$r$-matrix''.

As an example we will evaluate the metric in the case of GL${}_q(n)$.
The $\hat{r}$-matrix of GL${}_q(n)$ satisfies a characteristic equation
\begin{equation}
\hat{r}^2 - \lambda \hat{r} - 1 = 0
\end{equation}
which we can use in the form
\begin{equation}
r^{-1}_{21} = r_{12} - \lambda P_{12},
\end{equation}
where $P^{ij}{}_{kl} = \delta^i_l \delta^j_k$ is the permutation matrix,
to replace $(r^{-1}_{21})^{t_2}$ in equation (\ref{eta}).
That gives
\begin{equation}
\begin{array}{rcl}
\eta_{12} & = & - \left(P_{12}{}^{t_2} \left( (r_{12}{}^{t_2})^{-1}
\right) \right)^{t_2}\\
	& = & - \mbox{tr}_3\left(P_{23} (r_{23}{}^{t_3})^{-1} \right) P_{12}\\
	& = & - D_2 P_{12}.
\end{array}\label{etagln}
\end{equation}
In the last step we have used
\begin{equation}
D \equiv <u,t> = \mbox{tr}_2\left(P (r^{t_2})^{-1} \right),
\end{equation}
where $u \equiv \cdot (S \otimes i\!d) \R^{21}$ is the element of \U\
that implements the square of the antipode.
With the explicit formula ($\eta_{12} = - D_2 P_{12}$) for the metric
we immediately find an expression \cite{SWZ3} for the exterior
derivative $\dl$ on functions in terms of $X$ and the Maurer-Cartan form
$\Omega = t^{-1} \dl t$:
\begin{equation}
\dl = - \mbox{tr}( D^{-1} \Omega X). \z \mbox{\it (on functions)}
\end{equation}

The pure braid approach to the construction of quantum Lie algebras is
however particularly important in cases (like the 2-dim quantum euclidean
group)
where there is no quasitriangular Hopf algebra and where the $b_i$s
are not given by the elements of $t^i{}_j$.

\subsection{Various Types of Quantum Lie Algebras}

The functions $c^{j} := -S b^{j}$ play the role of coordinate functions.
Their span$\{c^{j}\} $ $=: R^{\perp }$ is the vector space dual to the quantum
tangent space $\tq$ = span$\{\chi_i\}$,
such that
\begin{equation}
\begin{array}{rcl}
1 \oplus \tq \oplus \tq^{\perp } & = & \U\\
1 \oplus R^{\perp } \oplus R & = & \A
\end{array}
\end{equation}
as vector spaces, with\footnote{We write here vector spaces in place of
their elements in an  obvious notation.}
\begin{equation}
<\tq , R> = 0,\z <\tq^{\perp },R^{\perp }> = 0.
\end{equation}
Let $\widetilde{R^{\perp }} =$span$\{b^{i}\}$ and $\widetilde{R}$ be the spaces
obtained from $R^{\perp }$ and $R$ by application of $S^{-1}$ to all their
elements.
In the following we will state various desirable properties that
different kinds of quantum Lie algebras might have; we will comment
on their significance and we will give the corresponding expressions
in the dual space. The proofs are given in the appendix.
\begin{equation}
i)\z\fbox{$\DA \tq \subset \tq \otimes \A \z\Leftrightarrow\z
\Delta ^{Ad} \widetilde{R} \subset \widetilde{R} \otimes \A$}
\end{equation}
The left hand side states the right invariance of $\tq$, which is
important for the covariance of the Cartan identity (\ref{CARTAN})
and the invariance of the realization (\ref{exder}) of $\dl$.
The right hand side is essential to Woronowicz's formulation
of the differential calculus because it allows to consistently
set $\omega _{\widetilde{R}} = 0$.
\begin{equation}
ii)\z\fbox{$ \Delta  \tq \subset \U \otimes (\tq \oplus 1) \z\Leftrightarrow\z
\A R = R$}
\end{equation}
The left hand side is necessary to ensure the existence of $f-\omega $
commutation relations that are consistent with an undeformed
Leibniz rule for $\dl$. It also implies a quadratic quantum commutator
for the $\chi _{i}$:
\begin{equation}
\chi _{k} \ad \chi _{l}
\equiv \Lix{k}(\chi _{l}) = \chi _{b} \chi _{c} (\delta ^{c}_{k}\delta ^{b}_{l}
-
\hat{R}^{cb}{}_{kl})
= \chi _{a} <\chi _{k},T^{a}{}_{l}> = \chi _{a} f_{k}{}^{a}{}_{l},
\end{equation}
where
\begin{equation}
\hat{R}^{cb}{}_{kl} = <O_{k}{}^{b},T^{c}{}_{l}>
\end{equation}
is the so-called ``big R-matrix''.
If $ii)$ is not satisfied we have the choice of giving
up the $f-\omega $ commutation relations, so that the algebra of forms $\Lambda
$
is only a left \A-module, or we can try a generalized Leibniz rule for $\dl$.
The right hand side of the equation is equivalent to $\widetilde{R} \A =
\widetilde{R}$
and states that $\widetilde{R}$ is a right
\A-ideal; it is the second fundamental ingredient
of Woronowicz's theory. If the Leibniz rule is satisfied then $ii)$ follows
from
$\omega_r = 0 \Rightarrow r \in \widetilde{R} \oplus 1$:
Let $a \in \A$, then
\begin{equation}
\omega_{r a} = S(a_{(1)}) S(r_{(1)}) \dl(r_{(2)} a_{(2)}) = S(a_{(1)}) \omega_r
a_{(2)}
+ \epsilon(r) \omega_a = 0,
\end{equation}
$\epsilon(r a) = \epsilon(r)\epsilon(a) = 0$ and hence $r a \in \widetilde{R}$.
$\A R = R$ is in agreement with the intuitive picture that
the ideal $R$ is spanned by polynomials in the $c^{i}$ of order 2 or higher,
$i.e.$ span$\{e_{i}\} \approx  \{1,c^{i},c^{i} c^{j}, \ldots\}$.
\begin{equation}
iii)\z\fbox{$ \Delta ^{Ad} \widetilde{R^{\perp }} \subset \widetilde{R^{\perp
}}
\otimes \A \z\Leftrightarrow\z
\DA \tq^{\perp } \subset \tq^{\perp } \otimes \A$}
\end{equation}
The right hand side keeps us out of trouble with covariance when
we set $\I_{{\cal T}_q^{\perp }} = 0$.
The left hand side is a {\em sufficient} condition for
$\DA(\tq^{*}) \subset \tq^{*} \otimes \A$. Quantum Lie algebras that satisfy
$iii)$ have particular nice properties in connection
with pure braid elements and a (Killing) metric.
That merits a special name for them:
\begin{quote}
Quantum Lie Algebra of {\bf type I\ }: $i)$,$ii)$,$iii)$\\
Quantum Lie Algebra of {\bf type II}: $i)$,$ii)$
\end{quote}
We will mainly be dealing with  type I, in fact, all examples of
quantum group calculi known to us are of this type.
Quantum Lie algebras of type II are mathematically equivalent
to Woronowicz's \cite{W2} theory.
\begin{equation}
iv)\z\fbox{$ \Delta  R^{\perp } \subset \A \otimes (R^{\perp } \oplus 1)
\z\Leftrightarrow\z
\U \tq^{\perp } = \tq^{\perp }$}
\end{equation}
The LHS enables us to define {\bf partial derivatives} instead of
left-invariant ones: It implies
$\Delta  c^{i} = M^{i}{}_{j} \otimes c^{j} + c^{i} \otimes 1$ with
$\Delta  M = M \dot{\otimes} M$, $S M = M^{-1}$, $\epsilon (M) = I$ and then
$\chi _{k} c^{i} = M^{i}{}_{k} + M^{i}{}_{j} <O_{k}{}^{l},c^{j}> \chi _{l} +
c^{i} \chi _{k}$,
such that $\partial _{n} := S^{-1}M^{k}{}_{n} \chi _{k}$ gives a commutation
relation
\begin{equation}
\partial _{n} c^{i} = \delta _{n}^{i} +
\left( S^{-1}M^{k}{}_{n} M^{i}{}_{j}
<O_{k}{}^{l},c^{j}> M^{m}{}_{l}
+ S^{-1}M^{k}{}_{n} c^{i} M^{m}{}_{k}\right) \partial _{m} \label{part}
\end{equation}
worthy of a partial derivative. (In the case of GL${}_q(n)$ we can use
(\ref{etagln}) to show
that $c^{(mn)} = (D^{-1})^n{}_k St^k{}_m$, $M^{(mn)}{}_{(ij)} = St^i{}_m
\delta^n_j$, and
$\partial_{(ij)} = t^i{}_k X^k{}_j$.) The exterior derivative (on functions)
becomes
\begin{equation}
\dl = \omega^i \chi_i = \dl(c^j) S^{-1}(M^i{}_j) M^n{}_i \partial_n = \dl(c^n)
\partial_n.
\end{equation}
\begin{equation}
v)\z\fbox{$ \Delta  R^{\perp } \subset (R^{\perp } \oplus 1) \otimes \A
\z\Leftrightarrow\z \tq^{\perp } \U = \tq^{\perp }$}
\end{equation}
This and $ii)$ imply quadratic $\chi -c$ commutation relations that
close in terms of the elements of $\tq$ and $R^{\perp }$.
The right hand sides of $iv)$ and $v)$ state that $\tq$ is a
left (right) \U-ideal, which supports the picture
of a Poincare-Birkhoff-deWitt type basis for \U\ in terms
of the $\chi _{i}$, $i.e.$ $\{1,\chi _{i},\chi _{i} \chi _{j},\ldots\}$.
Here and in the discussion following $ii)$ we have to be careful
though with higher order conditions on the generators.

\subsection{Universal Calculus}

Given (infinite) linear bases $\{ e_{i}\}$ of \U\ and
$\{f^{i}\}$ of \A\ we can always construct new counit-free elements
$\vec{e_{i}} := e_{i} - 1 \epsilon (e_{i})$  and $\vec{f^{i}} := f^{i} - 1
\epsilon (f^{i})$
that span (infinite) spaces $\tq^{u}$ and $R^{\perp u}$ respectively,
satisfying
properties $i)$ through $v)$; in fact $1 \oplus \tq^{u} = \U$ and
$1 \oplus R^{\perp u} = \A$ as vector spaces.
The $f-\omega $ commutation relations, however, become trivial in that they
are equivalent to the Leibniz rule for $\dg$;\footnote{To distinguish
this calculus from quantum Lie algebras we use the symbol $\dg$
instead of $\dl$ for the exterior derivative} we are hence dealing
with a Connes type calculus, a ``Universal Calculus on
Hopf Algebras''.  It is interesting to see what happens to
the formula for the partial derivatives in this limit:

\subsubsection{A Subbialgebra
and the Vacuum Projection
Operator}

To simplify notation we will assume that the infinite bases of \U\
and \A\ have been arranged in such a way that $e_{0} = 1_{{\cal U}}$,
$f^{0} = 1^{{\cal A}}$ and $e_{i}$, $f^{i}$ with
$\epsilon (e_{i}) = \epsilon (f^{i}) = 0$  for $i = 1,\ldots,\infty $ span
$\tq$ and $R^{\perp }$
respectively. Greek indices $\alpha ,\beta ,\ldots$ will run from
$0$ to $\infty $ whereas Roman indices $i,j,k,\ldots$ will only
take on values from $1$ to $\infty $ unless otherwise stated.
A short calculation gives
\begin{equation}
\Delta  f^{i} = M^{i}{}_{k} \otimes f^{k} + f^{i} \otimes 1,\z M^{i}{}_{k} =
f^{i}_{(1)}<e_{k},f^{i}_{(2)}>
\end{equation}
and
\begin{equation}
\Delta  M = M \dot{\otimes} M,\z S(M) = M^{-1},\z \epsilon (M) = I,
\end{equation}
in accordance with axiom $iv)$.
Using the definition from the previous section we will now write down
partial derivatives
\begin{equation}
\partial _{n} = S^{-1}(M^{l}{}_{n}) e_{l},\z (l\geq  1 !)
\end{equation}
which take on a peculiar form when using the explicit expression for $M$
\begin{equation}
\begin{array}{rcl}
\partial _{n}      & = & S^{-1}(f^{l}_{(1)})<e_{n},f^{l}_{(2)}> e_{l}\\
        & = & S^{-1}(f^{\alpha }_{(1)})<e_{n},f^{\alpha }_{(2)}> e_{\alpha }\\
        & = & S^{-1}(f^{\alpha }) <e_{n},f^{\beta }> e_{\alpha } e_{\beta } \\
        & = & S^{-1}(f^{\alpha }) e_{\alpha } e_{n}\\
        & = & E e_{n},
\end{array}
\end{equation}
where we have introduced the ``vacuum projector'' $E = S^{-1}(f^{\alpha })
e_{\alpha }$ in the last step.
$E$ \cite{CSW} has interesting properties like
\begin{eqnarray}
E a & = & E \epsilon (a), \z a \in \A,\\
x E & = & E \epsilon (x), \z x \in \U,\\
E^{2} & = & E.
\end{eqnarray}
Note also that
$E = \partial _{0} - 1$.
As expected we can express $\dg$ on functions in terms of partial
derivatives
\begin{equation}
\dg(f) = \dg(f^{i}) \partial _{i}(f).
\end{equation}
The partial derivatives are of course no longer left invariant,
but it turns out that we can actually define a coproduct for them
making the space $E \U = \{Ey;y \in \U\} \subset \smash$ a
unital bialgebra.
Inspired by
\begin{equation}
E y f = <y_{(1)},f> E y_{(2)} = (E y_{(1)})(f) E y_{(2)}
\end{equation}
we define
\begin{equation}
\Delta _{E}(E y) = E y_{(1)} \otimes E y_{(2)},\z \epsilon _{E}(E y) = \epsilon
(y),\z 1_{E} = E,
\end{equation}
in consistency with the axioms for a bialgebra.
$E \U$ is however not a Hopf algebra because it does not have an
antipode --- at least not with respect to the multiplication in $\smash$
--- so $E \U$ might be of use as an example of a quantum plane.

\subsubsection{Quantum Lie Algebras within a Universal Calculus}

If the span $\tq^{\mbox{\tiny univ.}}$ of the generators
$\{e_{a} | a = 1,\ldots,\infty \}$ of the universal calculus
contains a finite dimensional subspace, $\tq$ spanned by
$\{\chi _{i} | i = 1,\ldots,N\}$, that satisfies axioms
$i)$ and $ii)$ then one may ask how to obtain the finite calculus
from the infinite one. Let $\dg$ be the exterior derivative of the
universal calculus and $\dl$ the exterior derivative of the finite
calculus. One might be tempted to try an ansatz like
$\dg = \dl + \dl^{\perp }$,
where $\dg = \omega ^{a} e_{a}$ and $\dl = \omega ^{i} \chi _{i}$  on
functions. This equation
is covariant if axiom $iii)$ is also satisfied, but we run into
problems with the $f-\omega $ commutation relations. From the Leibniz rule
for $\dg$ we obtain
\begin{equation}
f \omega ^{i} = \omega ^{j} O_{j}{}^{i}(f) + \omega ^{r} \Theta
_{r}{}^{i}(f),\z i= 1,\ldots,N;\x r = N+1,\ldots,\infty,
\end{equation}
$i.e.$ the $f-\omega $ commutation relations do not close within the
finite calculus. So unless one decides to do without a bicovariant calculus
we have to make the second term vanish. The naive choice is to try
and set $\Theta $ equal to zero. This could be nicely expressed in terms
of another axiom
$$ \Delta  \tq^{\perp } \subset \U \otimes (\tq^{\perp } \oplus 1)
\z\Leftrightarrow\z
\A R^{\perp } = R^{\perp },$$
but the right hand side neither has a classical limit nor does it lend
itself to a description of $\A$ in terms of a Poincare-Birkhoff-deWitt
basis. The only choice left is to set the forms $\omega ^{r}$ corresponding
to functions in $R$ (recall: $<\tq,R> = 0$)
equal to zero. Following Connes' and Woronowicz's approach
we hence set
\begin{equation}
\omega _{R} = 0 \z\Rightarrow\z \dg  \to  \dl.
\end{equation}

\section[Calculus of Functions, Vector Fields and Forms]{Calculus
of Functions, Vector Fields and\\ Forms} \label{S:Cal}

The purpose of this section is to generalize the Cartan calculus of
ordinary {\em commutative} differential geometry to the case of quantum
Lie algebras.
As in the classical case, the Lie derivative of a function is
given by the action of the corresponding vector field, {\em i.e.}
\begin{equation}
\begin{array}{l}
\Li_{x}(a) = x \tr a =  a_{(1)} < x, a_{(2)} >,\\
\Li_{x} a = a_{(1)} < x_{(1)}, a_{(2)} > \Li_{x_{(2)}}.
\end{array} \label{XA}
\end{equation}
The action on products is given through the coproduct of $x$
\begin{equation}
x \tr a b = (x_{(1)} \tr a) (x_{(2)} \tr b). \label{XAB}
\end{equation}
The Lie derivative along $x$ of an element $y \in  \U$ is given by the
adjoint action in \U:
\begin{equation}
\Li_{x}(y) = x \ad y =  x_{(1)} y S(x_{(2)}).
\label{xady}
\end{equation}
To find the action of $\Ix{i}$ we can now attempt to use the Cartan
identity (\ref{CARTAN})\footnote{The idea is
to use this identity as long as it is consistent
and modify it if needed.}
\begin{equation}
\begin{array}{rcl}
\chi _{i} \tr a & = & \Lix{i}(a)\\
        & = & \Ix{i}(\dl a) + \dl(\Ix{i} a).
\end{array}
\end{equation}
As the inner derivation $\Ix{i}$ contracts 1-forms and is zero on
0-forms like $a$, we find
\begin{equation}
\Ix{i}(\dl a) = \chi _{i} \tr a = a_{(1)} < \chi _{i}, a_{(2)} >.
\label{incomplete}
\end{equation}
An equation like this could not be true for any $x \in \U$ because from the
Leibniz
rule for \dl\ we
have $\dl(1) = \dl(1\cdot  1) = \dl(1) 1 + 1 \dl(1) = 2 \dl(1)$
and any $\I_{x}$ that gives a non-zero result
upon contracting $\dl(1)$ will hence lead to a contradiction. From
(\ref{incomplete}) we see that the trouble makers would be  $x \in \U$
with $\epsilon (x) \neq 0$, but as $\epsilon (\chi _{i}) = 0$ we have nothing
to worry about.
Without loss of generality we can now set
\begin{equation}
\dl(1) \equiv 0\x\mbox{and}\x \I_{1} \equiv  0.
\end{equation}
Next consider for any form $\al$
\begin{equation}
\begin{array}{rcl}
\Lix{i}(\dl \al ) & = & \dl(\Ix{i} \dl \al )
                      + \Ix{i}(\dl \dl \al )\\
                & = & \dl(\Lix{i} \al ) + 0,
\end{array}
\label{Ld}
\end{equation}
which shows that Lie derivatives commute with the exterior
derivative; \ie\\  $\Lix{i} \dl = \dl  \Lix{i}$.\x We will later need to extend
this equation to all elements $x \in \U$:
\begin{equation}
\Li_{x} \dl = \dl  \Li_{x}.
\end{equation}
{}From this and (\ref{XA}) we find
\begin{equation}
\Li_{x} \dl(a) = \dl(a_{(1)}) < x_{(1)}, a_{(2)} > \Li_{x_{(2)}}.
\end{equation}
To find the complete commutation relations of $\Ix{i}$ with functions
and forms rather than just its action on them,
we next compute the action of $\Lix{i}$ on a product of functions
$a$, $b$ $\in \A$
\begin{equation}
\begin{array}{rcl}
\Lix{i}(a b) & = & \Ix{i} \dl(a b)\\
           & = & \Ix{i}(\dl(a) b + a \dl(b))
\end{array}
\end{equation}
and compare with equation (\ref{XAB}).
Recalling that the $\chi _{i}$  have coproducts of the form
\begin{equation}
\Delta \chi _{i} = \chi _{i} \otimes 1 + O_{i}{}^{j} \otimes \chi _{j},\z
O_{i}{}^{j} \in
\U,\label{DO}
\end{equation}
we obtain
\begin{equation}
\begin{array}{rcl}
\I_{\chi _{i}} a &=& (O_{i}{}^{j} \tr a) \;\I_{\chi _{j}}\\
&=& \Li_{O_{i}{}^{j}}(a) \;\I_{\chi _{j}},
\end{array}
\end{equation}
if we assume that the commutation relation of $\I_{\chi _{i}}$ with $\dl(a)$ is
of the general form
\begin{equation}
\I_{\chi _{i}} \dl(a) = \underbrace{\I_{\chi _{i}}(\dl a)}_{\in \A} +
\mbox{``braiding term''}\cdot \I_{\chi _{?}}\,.
\end{equation}
A calculation of $\Li_{\chi _{i}}(\dl(a) \dl(b))$ along similar lines
gives in fact
\begin{equation}
\begin{array}{rcl}
\I_{\chi _{i}} \dl(a) &=& (\chi _{i} \tr a) - \dl(O_{i}{}^{j} \tr a)
\;\I_{\chi _{j}}\\
&=&\I_{\chi _{i}}(\dl a) - \Li_{O_{i}{}^{j}}(\dl a) \;\I_{\chi _{j}}
\end{array}
\end{equation}
and we propose for any $p$-form $\al$:
\begin{equation}
\I_{\chi _{i}} \al = \I_{\chi _{i}}(\al) +  (-1)^{p} \Li_{O_{i}{}^{j}}(\al)\;
\I_{\chi _{j}}.
\end{equation}

Missing in our list are commutation relations of Lie derivatives with
vector fields and inner derivations.
The product in \U\ can be expressed
in terms of a right coaction $\DA: \U \rightarrow \U \otimes \A$,
denoted $\DA(y) = y^{(1)} \otimes y^{(2)'}$, such that
$x y = y^{(1)} <x_{(1)},y^{(2)'}> x_{(2)}$.
In the context of (\ref{xady}), this gives
\begin{eqnarray}
\Li_{x}(y) &=& x_{(1)} y S(x_{(2)}) = y^{(1)} <x,y^{(2)'}>,\\
\Li_{x} \Li_{y} &=& \Li_{\mbox{\small \pounds}_{x_{(1)}}(y)} \Li_{x_{(2)}} =
\Li_{y^{(1)}} <x_{(1)},y^{(2)'}> \Li_{x_{(2)}}.
\end{eqnarray}
For the special case $\chi _{i},\chi _{j} \in \tq$ that becomes
\begin{equation}
\begin{array}{rcl}
\Lix{i} \Lix{k} & = & \Lix{i}(\Lix{k}) + \Lio{i}{j}(\Lix{k})\Lix{j}\\
        & = & \Lix{l} f_{i}{}^{l}{}_{k}  + \Lix{a}  \Lix{b} \hat{R}^{ab}{}_{ik}
\end{array}
\end{equation}
and --- using the Cartan identity ---
\begin{equation}
\begin{array}{rcl}
\Lix{i} \Ix{k} & = & \Lix{i}(\Ix{k}) + \Lio{i}{j}(\Ix{k})\Ix{j}\\
               & = & \Ix{l} f_{i}{}^{l}{}_{k} + \Ix{a}  \Lix{b}
\hat{R}^{ab}{}_{ik},
\end{array}
\end{equation}
where
\begin{equation}
\hat{R}^{ab}{}_{ik} = <O_{i}{}^{b} , T^{a}{}_{k}>.
\end{equation}

\subsection{Maurer-Cartan Forms}

The most general left-invariant 1-form can be written \cite{W2}
\begin{equation}
\om_{b} := S(b_{(1)}) \dl(b_{(2)}) = - \dl(S b_{(1)} ) b_{(2)}
\end{equation}
\begin{equation}
(\mbox{\em left-invariance:}\x\AD(\om_{b}) =
S(b_{(2)}) b_{(3)} \otimes S(b_{(1)}) \dl(b_{(4)})
= 1 \otimes \om_{b} ),
\end{equation}
corresponding to a function $b \in \A$. If this function happens to
be $t^{i}{}_{k}$, where $t \in M_{m}(\A)$ is an $m \times m$ matrix
representation of \U\
with $\Delta (t^{i}{}_{k}) =$\mbox{$t^{i}{}_{j} \otimes t^{j}{}_{k}$} and
$S(t)=t^{-1}$, we obtain
the well-known Cartan-Maurer form $\om_{t} = t^{-1} \dl(t) =: \Omega $. Here is
a
nice formula for the exterior derivative of $\om_{b}$:
\begin{equation}
\begin{array}{rcl}
\dl(\om_{b}) & = & \dl(S b_{(1)} ) \dl(b_{(2)})\\
&=& \dl(S b_{(1)} ) b_{(2)} S(b_{(3)}) \dl(b_{(4)})\\
&=& - \om_{b_{(1)}} \om_{b_{(2)}}.
\end{array}
\end{equation}
The Lie derivative is
\begin{equation}
\begin{array}{rcl}
\Li_{\chi }(\om_{b})&=&\Li_{\chi _{(1)}}(S b_{(1)} ) \Li_{\chi _{(2)}}(\dl
b_{(2)}) \\
&=&<\chi _{(1)},S(b_{(1)})> S(b_{(2)}) \dl(b_{(3)}) <\chi _{(2)},b_{(4)}>\\
&=&\om_{b_{(2)}} <\chi ,S(b_{(1)})b_{(3)}> \\
&=&<\chi _{(1)},S(b_{(1)})> \om_{b_{(2)}} <\chi _{(2)},b_{(3)}>.
\end{array} \label{XOM}
\end{equation}
For $\chi  = \chi _{i}$ and $b = t^{k}{}_{n}$ this becomes a quantum
commutator:
\begin{equation}
\begin{array}{rcl}
\Lix{i}(t) & = & <\chi _{i},S t>\cdot \Omega  +
        <O_{i}{}^{j},S t>\cdot \Omega \cdot <S^{-1} \chi _{j} ,S t>\\
& = & <\chi _{i},S t>\cdot \Omega  -
        <O_{i}{}^{j},S t>\cdot \Omega \cdot <S^{-1}O_{j}{}^{k} ,S t>\cdot <\chi
_{k},S t>\\
& = & <\chi _{i},S t>\cdot \Omega  -
        \Lio{i}{k}(\Omega )\cdot <\chi _{k},S t>
\end{array}
\end{equation}
and, if we denote a $S t$-matrix representation for the moment by
``$\widetilde{\x}$'',
\begin{equation}
\Lix{{}}(t) = \tilde{\chi }\cdot t - \tilde{O}\cdot t\cdot \tilde{O}^{-1}\cdot
\tilde{\chi }
=: \left[ \tilde{\chi } , t \right]_{q}.
\end{equation}
The contraction of left-invariant forms with $\Ix{{}}$
--- {\em i.e.} by a {\em left-invariant} $x \in \U$ ---
gives a number in the field $k$ rather than a function
in \A\ as was the case for $\dl(a)$. (The result must be a number because
the only invariant function is 1.)
\begin{equation}
\begin{array}{rcl}
\Ix{{}}(\om_{b}) & = & \Ix{{}}(- \dl(S b_{(1)} ) b_{(2)})\\
&=& - \Ix{{}}(\dl S b_{(1)} ) b_{(2)}\\
& = & -<\chi ,S(b_{(1)})> S(b_{(2)}) b_{(3)}\\
&=& -<\chi ,S(b)>.
\end{array} \label{IOM}
\end{equation}
As an exercise and to check consistency we will compute the
same expression in a different way:
\begin{equation}
\begin{array}{rcl}
\Ix{i}(\om_{b}) & = & \Ix{i}(S b_{(1)} \dl(b_{(2)}))\\
& =&<O_{i}{}^{j},S(b_{(1)})> S(b_{(2)}) \Ix{j}(\dl b_{(2)})\\
& = & <O_{i}{}^{j},S(b_{(1)})> S(b_{(2)}) b_{(3)} <\chi _{j},b_{(4)}>\\
& =& <O_{i}{}^{j},S(b_{(1)})><\chi _{j},b_{(2)}>\\
&=&  -<\chi _{i},S(b)>.
\end{array}
\end{equation}

\subsubsection{The Exterior Derivative on Functions}

We would like to express the exterior derivative of a function $f$
in terms of the basis of 1-forms $\{\omega ^{i}\}$ with functional
coefficients.
There are two natural ans\"atze: $\dl(f) = \omega ^{j} a_{j}$  and $\dl(f) =
b_{j} \omega ^{j}$
with appropriate $a_{j}, b_{j} \in \A$. Applying the Cartan identity
(\ref{CARTAN}) to $f$ we find
$$ \Lix{i}(f) = a_{i} = \Lio{i}{j}(b_{j}),$$
giving two alternate expressions for $\dl(f)$ :
\begin{equation}
\dl(f) = \omega ^{j} \Lix{j}(f) = - \Li_{S \chi _{j}}(f) \omega ^{j}.
\end{equation}
The Woronowicz and Castellani groups use the second expression, while
we prefer the first one because it allows us to write $\dl$ as
an operator $\omega ^{j} \chi _{j}$ on \A. An operator expression just like
this,
but written in terms of partial derivatives, is at least classically valid
on all forms. (For quantum planes that also holds).
Combining the two expressions for $\dl$ one easily derives the
well-known $f-\omega $ commutation relations
\begin{equation}
f \omega ^{i} = \omega ^{j} \Lio{j}{i}(f).
\end{equation}
The classical limit is given by
$O_{j}{}^{i} \rightarrow 1 \delta ^{i}_{j}$, so that  forms
commute with functions.

\subsubsection{On the Invariance of $\dl = \omega _{b^{j}} \chi _{j}$}

Recall: $\DA(\om^{i}) = \om_{b^{i}_{(2)}} \otimes
S(b^{i}_{(1)}) b^{i}_{(3)} = -\om^{j} \otimes <S\chi _{j},b^{i}_{(2)}>
Sb^{i}_{(1)} b^{i}_{(3)}$. Assuming
$\DA \chi _{i} = \chi _{j} \otimes T^{j}{}_{i}$\x(axiom $i)\,$)  we would like
to show
\begin{equation}
\DA(\omega _{b^{i}} \chi _{i}) = \om_{b^{i}_{(2)}} \chi _{i}^{(1)} \otimes
S(b^{i}_{(1)}) b^{i}_{(3)} \chi _{i}^{(2)'} =
\omega ^{i} \chi _{i} \otimes 1,
\end{equation}
$i.e.$
\begin{equation}
\DA(\om^{i}) = \om^{j} \otimes S^{-1}(T^{i}{}_{j}),
\end{equation}
or equivalently
\begin{equation}
-<S \chi _{k},b^{i}_{(2)}> S(b^{i}_{(1)}) b^{i}_{(3)}  = -S^{-1}(<\chi
_{k}^{(1)} , S b^{i}> \chi _{k}^{(2)'}).
\end{equation}
This turns out to be a purely Hopf algebraic identity for {\em any}
$x \in  \U,\x a \in \A$ (see equation \ref{alid}):
\begin{equation}
S^{-1}(x^{(2)'})<x^{(1)},S a> =  <x , S a_{(2)}> S a_{(1)} a_{(3)}.
\end{equation}
We see that axiom $iii)$ is not {\em necessary} for the invariance of $\dl$,
it is however sufficient.

\subsection{Tensor Product Realization of the Wedge}

As in the classical theory we make an ansatz for
the product of two forms in terms of tensor products
\begin{equation}
\om^{i} \wedge \om^{j} = \om^{i} \otimes \om^{j} - \hat{\sigma }^{ij}{}_{mn}
\om^{m} \otimes \om^{n},
\end{equation}
with as yet unknown numerical constants $\hat{\sigma }^{ij}{}_{mn} \in
k$, and define $\I_{\chi _{i}}$ to act on this product by contracting
in the first tensor product space, {\em i.e.}
\begin{equation}
\I_{\chi _{i}}(\om^{j} \wedge \om^{k}) = \delta _{i}^{j} \om^{k} -
\hat{\sigma }^{jk}{}_{mn} \delta _{i}^{m} \om^{n}.
\end{equation}
For higher order forms we can use an operator $W$  that recursively translates
wedge products into the tensor product representation:
\begin{equation}
\begin{array}{l}
W: \Lambda ^{p}_{q} \rightarrow
{\cal T}^{*}_{q} \otimes \Lambda ^{p-1}_{q},\x p \geq 1,\\
W(\alpha ) = \om^{n} \otimes \I_{\chi _{n}}(\alpha ),
\end{array}
\end{equation}
for any p-form $\alpha $. Two examples:
\begin{equation}\label{twof}
\begin{array}{rcl}
\om^{j} \wedge \om^{k}
        & = & \om^{n} \otimes \I_{\chi _{n}}(\om^{j} \wedge \om^{k})\\
        & = & \om^{n} \otimes
                (\delta ^{j}_{n} \om^{k} - \Li_{O_{n}{}^{m}}(\om^{j})
                \delta ^{k}_{m})\\
        & = & \om^{j} \otimes \om^{k} - \om^{n} \otimes
                \Li_{O_{n}{}^{k}}(\om^{j})\\
        & = & \om^{j} \otimes \om^{k} - \om^{n} \otimes \om^{m}
                \hat{\sigma }^{jk}{}_{nm}
\end{array}
\end{equation}
and, after a little longer computation that uses $W$ twice,
\begin{equation}
\begin{array}{rcl}
\om^{a} \wedge \om^{b} \wedge \om^{c}
        & = & \om^{a} \otimes (\om^{b} \wedge \om^{c})
                - \om^{i} \otimes (\om^{j} \wedge \om^{c}) \hat{\sigma
}^{ab}{}_{ij}\\
        &  & + \om^{i} \otimes (\om^{j} \wedge \om^{k}) \hat{\sigma
}^{al}{}_{ij}
                \hat{\sigma }^{bc}{}_{lk}\\
        & = & \om^{a} \otimes \om^{b} \otimes \om^{c}
                - \om^{a} \otimes \om^{j} \otimes \om^{k} \hat{\sigma
}^{bc}{}_{jk}\\
        &   & -\om^{i} \otimes \om^{j} \otimes \om^{c} \hat{\sigma
}^{ab}{}_{ij}
                + \om^{i} \otimes \om^{j} \otimes \om^{k}
                \hat{\sigma }^{lc}{}_{jk} \hat{\sigma }^{ab}{}_{il}\\
        &   & + \om^{i} \otimes \om^{j} \otimes \om^{k}
                \hat{\sigma }^{al}{}_{ij} \hat{\sigma }^{bc}{}_{lk}
                - \om^{i} \otimes \om^{j} \otimes \om^{k}
                \hat{\sigma }^{an}{}_{il} \hat{\sigma }^{bc}{}_{nm}
\hat{\sigma }^{lm}{}_{jk},
\end{array}
\end{equation}
where $\hat{\sigma }^{ij}{}_{mn} = <O_{m}{}^{j},S^{-1}(T^{i}{}_{n})>$.
In some cases this expression can be further simplified with the help of
the characteristic equation of $\hat{\sigma }$.

Equation (\ref{twof}) can be used to obtain (anti)commutation relations
between the $\om^{i}$s; by using the characteristic equation for
$\hat{\sigma }$, projection matrices orthogonal to the antisymmetrizer $I-
\hat{\sigma }$ can be found, and these will annihilate $\om^{i} \wedge
\om^{j}$.  The resulting equations will determine how to commute the
1-forms. In some rare cases the $\omega -\omega $ commutation relations
are of higher than second order. We are then forced to consider
orthogonal projectors to the operator $W$, introduced below.
There is another reason why we want to emphasize the tensor product
realization of the wedge product rather than commutation relations
given in terms of projection operators:
In the case of quantum groups in the A, B, C and D series
$\hat{\sigma }$ typically has one eigenvalue equal to 1, so there is exactly
one projection operator $P_{0}$  orthogonal to $(1 - \hat{\sigma })$,
but while $(1 - \hat{\sigma })$ has a sensible classical limit ---
it becomes $(1 - P)$ where $P$ is the permutation matrix --- $P_{0}$, on the
other hand
might change discontinuously as q reaches 1 if $(1-\hat{\sigma })$ had other
eigenvalues $\lambda _{i}$ that become equal to 1 in that limit because
the corresponding projection operators $P_{i}$ will now {\em all} be orthogonal
to $(1 - P) = \left.(1-\hat{\sigma })\right|_{q=1}$. The approach of the group
in M\"unchen trying to circumvent this problem in the case of SO${}_{q}(3)$
was to impose additional conditions on the wedge product ``by hand'',
requiring that all projection operators $P_{i}$  (see above) vanish on it.
In the present context we would have to simultaneously
impose similar conditions on products of inner derivations {\em and}
check consistency of the resulting equations on  a case by case basis.

\subsubsection{Example: Maurer-Cartan-Equation}

\begin{equation}
\begin{array}{rcl}
\dl \omega ^{j}  & = & \dl \omega _{b^{j}} = -\omega _{b^{j}_{(1)}} \wedge
\omega _
{b^{j}_{(2)}}\\
        & = & -\omega _{S^{-1}(Sb^{j}_{(1)} b^{j}_{(3)})} \otimes \omega
_{b^{j}_{(
2)}}\\
        & = & -\omega ^{k} \otimes
                \omega ^{l} <-S\chi _{k},S^{-1}(Sb^{j}_{(1)}
b^{j}_{(3)})><-S\chi _
{l},b^{j}_{(2)}>\\
        & = & -\omega ^{k} \otimes \omega ^{l}
<\underbrace{(S^{-1}\chi _{k})_{(1)} \chi _{l}
                S(S^{-1}\chi _{k})_{(2)}}_{S^{-1}\chi _{k}
\ad \chi _{l}},Sb^{j}> \\
        & = & -f'_{k}{}^{j}{}_{l} \omega ^{k} \otimes \omega ^{l}.
\end{array}
\end{equation}
In the previous equation we have introduced the adjoint action of
a left-invariant vector field on another vector field. A short
calculation gives
\begin{equation}
S^{-1}\chi _{k} \ad \chi _{l}
= \chi _{b} \chi _{c} (\delta ^{c}_{k}\delta ^{b}_{l} - \hat{\sigma
}^{cb}{}_{kl})
= \chi _{a} <S^{-1}\chi _{k},T^{a}{}_{l}>=  \chi _{a} f'{}_{k}{}^{a}{}_{l}
\end{equation}
as compared to
\begin{equation}
\chi _{k} \ad \chi _{l}
\equiv \Li_{\chi _{k}}(\chi _{l}) = \chi _{b} \chi _{c} (\delta ^{c}_{k}\delta
^{b}
_{l} -
\hat{R}^{cb}{}_{kl})
= \chi _{a} <\chi _{k},T^{a}{}_{l}> = \chi _{a} f_{k}{}^{a}{}_{l},
\end{equation}
with
$\hat{R}^{cb}{}_{kl} = <O_{k}{}^{b},T^{c}{}_{l}>$.
The two sets of structure constants are related by
\begin{equation}
f_{k}{}^{a}{}_{l} = -f'_{i}{}^{a}{}_{l} R^{ij}{}_{kl}.
\end{equation}
Please see \cite{CaMo}
for a detailed discussion of such structure constants.

\subsection{Summary of Relations in the Cartan Calculus} \label{S:SoRCC}

\subsubsection{Commutation Relations}

For any $p$-form $\alpha $:
\begin{eqnarray}
\dl \alpha   & = & \dl(\alpha ) + (-1)^{p} \alpha  \dl\\
\I_{\chi _{i}} \alpha  & = & \I_{\chi _{i}}(\alpha ) + (-1)^{p}
\Li_{O_{i}{}^{j}}(\alpha ) \I_{\chi _{j}}\\
\Li_{\chi _{i}} \alpha  & = & \Li_{\chi _{i}}(\alpha ) +
\Li_{O_{i}{}^{j}}(\alpha ) \Li_{\chi _{j}} \label{il1}
\end{eqnarray}

\subsubsection{Actions}

For any function $f \in \A$, 1-form $\omega _{f} \equiv Sf_{(1)} \dl f_{(2)}$
and
vector field $\phi  \in \A\cross\U$:
\begin{eqnarray}
\Ix{i}(f) & = & 0\\
\Ix{i}(\dl f) & = & \dl f_{(1)}<\chi _{i},f_{(2)}>\\
\Ix{i}(\omega _{f}) & = & -<\chi _{i},S f>\\
\Lix{{}}(f) & = & \chi (f) = f_{(1)}<\chi ,f_{(2)}>\\
\Lix{{}}(\omega _{f})& = & \omega _{f_{(2)}} <\chi ,S(f_{(1)}) f_{(3)}>\\
\Lix{{}}(\phi )& = & \chi _{(1)} \phi  S(\chi _{(2)}) \label{il2}
\end{eqnarray}

\subsubsection{Graded Quantum Lie Algebra of the Cartan Generators}

\begin{eqnarray}
\dl \dl & = & 0\\
\dl \Lix{{}} & = & \Lix{{}} \dl\\
\Li_{\chi _{i}} & = & \dl \I_{\chi _{i}} + \I_{\chi _{i}} \dl\\
\left[\Lix{i},\Lix{k} \right]_{q} & = & \Lix{l} f_{i}{}^{l}{}_{k}\\
\left[\Lix{i},\Ix{k} \right]_{q} & = & \Ix{l} f_{i}{}^{l}{}_{k}
\end{eqnarray}
The quantum commutator $[\, , \,]_{q}$ is here defined as follows
\begin{equation}
\left[\Lix{i}, \Box \right]_{q} :=
\Lix{i} \Box - \Lio{i}{j}(\Box) \Lix{j}.
\end{equation}
This quantum Lie algebra becomes infinite dimensional as soon as we
introduce derivatives along general vector fields.

\subsection{Braided Cartan Calculus}

There are several graphical representations of the relations
that we derived in the previous sections. One that
emphasizes the nature of differential operators is
illustrated here at the example of equation (\ref{il1}):\\
\begin{center}
\input{[schupp.thesis]braid2.pic}
\end{center}
There is another graphical representation that is special in
as it shows
that we are in fact dealing with a graded and braided
Lie algebra in the sense of \cite{Md4}. Recall that in the braided
setting the coproducts and antipodes of the
generators $\{\chi _{i}\}$ take on the
classical linear form
\begin{equation}
\Delta  \chi _{i} = \chi _{i} \otimes 1 + 1 \otimes \chi _{i},\z S \chi _{i} =
- \chi _{i}\z\mbox{\it (braided)},
\end{equation}
while the multiplication  of tensor products acquires braiding
$$
(a \otimes b) \cdot  (c \otimes d) = a \Psi (b \otimes c) d \x\in W \otimes V,
$$
described by a ``braided-transposition''\cite{Md4}  operator
$\Psi _{V,W}:V \otimes W \to  W \otimes V$. This notation suggests that
the braiding is of a symmetric nature with respect to the two spaces $V$
and $W$. In the present case it turns out to be more fruitful to
assign all braiding to the generators $\chi _{i}$ --- or linear combinations
of them --- as they move through various objects. The general
braiding rule can be stated symbolically as
\begin{equation}
\Psi :\x \chi _{i} \otimes \Box \mapsto \Lio{i}{j}(\Box) \otimes \chi _{j},
\end{equation}
where $\chi _{i}$  could be part of an object like $\Li$ or $\I$. If
$\chi _{i}$ is part of $\I$, $i.e.$  of degree -1, there will be an
additional $(-1)^{p}$ grading, depending on the degree $p$ of $\Box$.
Here is a summary of all braid relations involving Cartan generators:
For $\Box \in \{\Lix{k},\Ix{k},\dl,\mbox{vector
fields},\mbox{forms},\mbox{functions}\}$
\begin{equation}
\Psi :\, \Lix{i} \otimes \Box  \mapsto  \Lio{i}{j}(\Box) \otimes \Lix{j},
\end{equation}
for $\Box \in \{\dl,\mbox{vector
fields},\mbox{forms},\mbox{functions}\}$
\begin{equation}
\Psi :\, \Ix{i} \otimes \Box  \mapsto  (-1)^{p} \Lio{i}{j}(\Box) \otimes
\Ix{j},
\end{equation}
and finally
\begin{equation}
\Psi :\, \dl \otimes \dl \mapsto - \dl \otimes \dl.
\end{equation}
Let us now look at the graphical representation of
the adjoint action (\ref{il2})
$(\chi _{i},\phi ) \mapsto \Lix{i}(\phi ) = \chi _{i(1)} \phi  S(\chi
_{i(2)})$:\\
\begin{center}
\input{[schupp.thesis]braid1.pic}
\end{center}
(In the right column we have translated the various graphical
manipulations into their algebraic counterparts.)
Taking this diagram as the definition of a braided (and graded) commutator
we can now express all Cartan relations in graphical form:\\
\paragraph{Lie derivatives.} Note that $\Lio{i}{j}(\dl) = \delta _{i}^{j} \dl$
because $\dl$ is invariant.\\
\input{[schupp.thesis]braid3.pic}\\
The relation $\left[\Lix{i},\Lix{k}\right]_{q} = \Lix{l} f_{i}{}^{l}{}_{k}$ has
a
very similar picture, so we did not show it here.
\paragraph{Inner derivations.} $\alpha $ is a $p$-form here.\\
\input{[schupp.thesis]braid4.pic}\\
\paragraph{Exterior derivative.} Here we use that $\dl$ is a derivation
in the sense $``\Delta(\dl)" = \dl \otimes 1 + 1 \otimes \dl$.\\
\input{[schupp.thesis]braid5.pic}\\
{\em Remark:} We have not given the full braided structure of the Cartan
Calculus here, but just the aspects of it that are of relevance to the
braided commutators. For the braided structure of {\it e.g.} the function
Hopf algebra, see \cite{Md4}.

\subsection{Lie Derivatives Along General Vector Fields}

So far we have focused on Lie derivatives and inner derivations along
{\em left-invariant} vector fields, {\em i.e.}
along elements of $\tq$. The classical theory
allows functional coefficients, {\em i.e.} the
vector fields need not be left-invariant.
Here we may introduce derivatives along
elements in the
$\A\cross\tq$ plane by the following set of equations
valid on forms:
(note: $\epsilon (\chi ) = 0$ for $\chi \in {\cal T}_{q}$)
\begin{eqnarray}
\I_{f \chi } & = & f \I_{\chi },\\
\Li_{f \chi } & = & \dl  \I_{f \chi } + \I_{f \chi } \dl,\\
\Li_{f \chi } & = & f \Li_{\chi } + \dl(f) \I_{\chi }
\label{liefx},\\
\Li_{f \chi } \dl & = & \dl \Li_{f \chi }.
\end{eqnarray}
Equation (\ref{liefx})
can be used to define Lie derivatives recursively on any form.
There does not seem to be a way to generalize the classical Lie
bracket to
include Lie derivatives of {\em vector fields} along {\em
arbitrary} elements of $\A\cross\U$ or $\A\cross{\cal T}_{q}$
in the quantum case. Exceptions are the right-invariant vector fields
$\widehat{x} \in \A\cross\U$, where
\begin{equation}
\Li_{\widehat{x}}(\phi ) = \widehat{x_{(1)}} \phi
\widehat{S^{^{-1}}x_{(2)}},\z \mbox{for }
\phi  \in \A\cross\U.
\end{equation}

\section{Quantum Planes Revisited}

With the new tools that we have developed in the previous sections
we are now ready to take a second look at quantum planes.
The first two sections that follow will be devoted to the
realization and action of quantum Lie algebra generators
on a quantum plane. After introducing the basic equations we will
spend some time on the important question of their covariance.
The third section finally gives an introduction to the construction of a
Cartan calculus on quantum planes with the surprising result
that the $\Li_{\partial }-x$ commutation must contain
inner derivation terms in order
to be consistent with a Lie derivative that commutes with $\dl$.
For notational simplicity we will however suppress these inner derivation terms
in the following two sections.

\subsection{Induced Calculus} \label{indcal}

In this section we wish to show how the calculus of the symmetry quantum
group induces a calculus on the plane.
We will study realizations of
quantum group generators in terms of the calculus on a quantum
plane. This will also give an explanation for the appearance
of ``inner derivation terms'' in the generalized product rule.

The central idea of this section is to
give the coordinate functions on the quantum plane functional
coefficients in \A, \ie\ to make them variable with respect to the
action of vector fields in \U. Let $x_{0}^{i} \in \fum$ be the ``fixed''
coordinate functions and define new variable ones via
$x^{i} := (t^{-1})^{i}{}_{j} x_{0}^{j}$. Instead of the
differentials $\dl x_{0}^{i}$ we will
use $\dg x^{i} = -(\Omega  x)^{i}$ because
\begin{equation}
\dg x = \dg t^{-1}\cdot  x_{0}
= t^{-1}\cdot t\cdot \dg(t^{-1})\cdot x_{0} = -t^{-1}\cdot \dg(t)\cdot
t^{-1}\cdot x_{0} = -\Omega \cdot x,
\end{equation}where $\dg$ is the exterior derivative on the
quantum group and $\Omega  = t^{-1} \dg t$ is the Maurer-Cartan Matrix.
By ``pullback'' the group derivative will become the derivative on
the plane, inducing a differential calculus there.
It then immediately follows that $\DA(\dl x^{i}) = \dl x^{j} \otimes
(t^{-1})^{j}{}_{i}$, which will ultimately give us the desired
commutativity between
Lie derivatives and $\dl$.

Turn now to the quantum group. Reserving Latin indices $i,j,\ldots$ for
the plane coordinates, let us use Greek indices for the adjoint
representation of the quantum group. Let $\{v_{\alpha }\}$
be a basis of
bicovariant generators with coproduct
$\Delta  v_{\alpha } = v^{\alpha } \otimes 1 + O_{\alpha }{}^{\beta } \otimes
v_{\beta }$ spanning $\tq \subset \U$
and let $\{\omega _{\alpha }\}$ be the dual basis of 1-forms; $\I_{v_{\alpha
}}(\omega ^{\beta }) = \delta _{\alpha }^{\beta }$,\x
$\Omega ^{i}{}_{j} =
\omega ^{\alpha } \I_{v_{\alpha }}(\Omega ^{i}{}_{j})
= -\omega ^{\alpha } <v_{\alpha },(t^{-1})^{i}{}_{j}>$. Via the Cartan identity
$\Li_{v} = \I_{v} \dg + \dg \I_{v}$ one computes actions of $\tq$ on $\fum$:
\begin{equation}
v_{\alpha } \tr x^{i}
= \I_{v_{\alpha }}(\dg x^{i}) = <v_{\alpha },(t^{-1})^{i}{}_{j}> x^{j}.
\end{equation}
Now we can make an ansatz for a realization of the
group generators in terms of
functions and derivatives on the plane\footnote{``$\doteq$" means: ``equal
when evaluated on $\fum$''}
\begin{equation}
v_{\alpha } \doteq J_{\alpha }^{i} \partial _{i},
\end{equation}
where $J_{\alpha }^{i} \in
\fum$ is easily computed, using $\partial _{i}(x^{j}) = \delta _{i}^{j}$, to be
\begin{equation}
J_{\alpha }^{i} = v_{\alpha }(x^{i}) = <v_{\alpha },(t^{-1})^{i}{}_{j}> x^{j}.
\end{equation}
In some lucky cases there is an inverse expression for the
partial derivatives on the plane in terms of the group generators.
With $\tilde{J_{i}^{\alpha }} \in \fum$
\begin{equation}
\partial _{i} = \tilde{J_{i}^{\alpha }} \otimes v_{\alpha }\tr,
\end{equation}
an expression that is classically only valid locally  and may exclude
some points unless we are dealing with an inhomogeneous group, but will
give explicit $\partial -x$ commutation relations if it exists:
\begin{equation}
\partial _{i} x^{j} =
\tilde{J_{i}^{\alpha }} v_{\alpha } x^{j} = \partial _{i}(x^{j}) +
\underbrace{\tilde{J_{i}^{\alpha }} O_{\alpha }{}^{\beta }(x^{j})
J_{\beta }{}^{k}}_{L_{i}{}^{k}(x^{j})} \partial _{k}. \label{nicedx}
\end{equation}

\subsubsection{Example: GL${}_{\frac {1}{q}}(2)$,
Manin-Wess-Zumino Quantum Plane}

The coordinate functions $x,y$ of the Manin plane satisfy commutation
relations $x y = q y x$ that are covariant under coactions of the
quantum matrix group GL${}_{\frac {1}{q}}(2)$.
This quantum group has four bicovariant
generators $v_{1},v_{2},v_{+},v_{-}$;
we will focus on the last two for the moment,
giving their fundamental $t^{-1}$ representations
\begin{equation}
<v_{+},t^{-1}> = \left(\begin{array}{cc}0 & q^{3}\\0 & 0\end{array}\right),\z
<v_{-},t^{-1}> = \left(\begin{array}{cc}0 & 0\\q & 0\end{array}\right)
\end{equation}
and the first tensor product representations
\begin{equation}
<\Delta  v_{+},t_{1}^{-1} \otimes t_{2}^{-1}>
= \left(\begin{array}{cccc}0&q^{4}&q^{3}&0\\
0&0&0&q^{5}\\0&0&0&q^{4}\\0&0&0&0\end{array}\right),
\end{equation}
\begin{equation}
<\Delta  v_{-},t_{1}^{-1} \otimes t_{2}^{-1}>
= \left(\begin{array}{cccc}0&0&0&0\\
q^{2}&0&0&0\\q&0&0&0\\0&q^{3}&q^{2}&0\end{array}\right).
\end{equation}
All these were obtained from
\begin{equation}
r_{\frac {1}{q}} = \left(\begin{array}{cccc}\frac {1}{q}
&0&0&0\\0&1&0&0\\0&-\lambda &1&0\\
0&0&0&\frac {1}{q}\end{array}\right).
\end{equation}
We immediately find
\begin{equation}
\partial _{x} = \tilde{J}_{x}^{\alpha }
v_{\alpha } =q^{-3} y^{-1} v_{+},\z \partial _{y} = q^{-1} x^{-1} v_{-},
\end{equation}
which we only have to check on pairs of functions because of the
form of (\ref{nicedx}):
\begin{equation}
\partial _{x} \left(\begin{array}{c}x x\\x y\\y x\\y y\end{array}\right) =
\left(\begin{array}{c}(1+q^{2}) x\\q^{2} y\\q y\\0\end{array}\right),\z
\partial _{y} \left(\begin{array}{c}x x\\x y\\y x\\y y\end{array}\right) =
\left(\begin{array}{c}0\\q x\\x\\y+q^{2} y\end{array}\right).
\end{equation}
{}From this we read off the following $\partial -x$ commutation relations in
perfect agreement with the results given in \cite{WZ}
\begin{eqnarray}
\partial _{x} x & =
& 1 + q^{2} x \partial _{x} + (q^{2}-1) y \partial _{y},\\
\partial _{x} y & = & q y \partial _{x},\\
\partial _{y} x & = & q x \partial _{y},\\
\partial _{y} y & = & 1 + q^{2} y \partial _{y}.
\end{eqnarray}
Using the other two generators $v_{1},v_{2}$ gives identical results.
This method works for all linear quantum planes and
can be formulated abstractly in terms of $r$-matrices.
The results for GL${}_q(n)$ are given in section~\ref{S:ndcc}.

If one does not want to extend the algebra by introducing inverses
$y^{-1},x^{-1}$ of the coordinate functions, it is also possible
to obtain the above commutation relations as a vanishing ideal
of $x y$ thereby also avoiding the questionable use of $\tilde{J}$.

\subsection{Covariance}

Let us collect some of the equations valid on a quantum plane.
Let $f,g \in \fum$ be functions
and $\partial _{i}$ be  derivatives on the quantum
plane, let $v_{a}$ be generators of the quantum Lie algebra
--- corresponding to the symmetry quantum group of the plane ---
with coproduct $\Delta  v_{a} = v_{a}
\otimes 1 + O_{a}{}^{b} \otimes v_{b}$, and let
$L_{i}{}^{j}$ be a linear automorphism of $\fum$:
\footnotetext{Careful: An expression linear in the partials may not
always exist, in particular for $e_{q}(2)$ we get a power series
instead. It {\em does} exist for Wess-Zumino type quantum planes
and then we have $J_{a}^{i} = <v_{a},(t^{-1})^{i}{}_{j}> x^{j}$.}
\addtocounter{footnote}{-1}
\begin{eqnarray}
v_{a} f & = & v_{a}(f) + O_{a}{}^{b}(f) v_{b},\\
v_{a} & \doteq & J_{a}^{i} \partial _{i},\footnotemark\\
\partial _{i} f & = & \partial _{i}(f) + L_{i}{}^{j}(f) \partial _{j}.
\end{eqnarray}
\addtocounter{footnote}{+1}
{}From this equations we can form a new one
\begin{equation}
\fbox{$J_{a}^{i} L_{i}{}^{k}(f) = O_{a}{}^{b}(f) J_{b}^{k}$},\label{newone}
\end{equation}
that can sometimes be rewritten as
\begin{equation}
L_{i}{}^{k}(f) = \tilde{J}_{i}^{a} O_{a}{}^{b}(f) J_{b}^{k}.
\end{equation}

\subsubsection{Examples}

\paragraph{Partial derivatives for a quantum group: }
For quantum groups that satisfy axiom $iv)$ we can construct partial
derivatives
from the left-invariant generators: $\partial_i = S^{-1}(M^a{}_i) v_a$. We can
then ``represent'' the generators $v_a$ in terms of these partial derivatives
by choosing $J_a^i = M^i{}_a$. Plugging this into equation (\ref{newone}) gives
$L_j{}^k(f) = S^{-1}(M^a{}_j) O_a{}^b(f) M^k{}_b$, in agreement with
(\ref{part}).

\paragraph{Quantum group as a quantum plane \protect\cite{Z3}: $\fum := \A$.}
\begin{quote}
Left-Invariant Generators: $\partial _{i} := v_{i}
\Rightarrow$ $J_{i}^{j} = \delta _{i}^{j},\x L_{i}{}^{j} =
O_{i}{}^{j}.$
\end{quote}
\begin{quote}
Plane-Like Generators: $\partial _{(ij)}:= t^{i}{}_{k} X^{k}{}_{j} \Rightarrow$
$J_{(kl)}^{(ij)} = (t^{-1})^{k}{}_{i} \delta _{l}^{j},$ \\
$L_{(lj)}{}^{(nm)}(f) = t^{l}{}_{i} O_{(ij)}{}^{(km)}(f) (t^{-1})^{k}{}_{n}.$
\end{quote}
\paragraph{Linear quantum plane:} The algebra of functions on the
linear quantum plane is invariant under coactions
of GL${}_{q}(N)$; $\DA(x^{i}) = x^{j} \otimes S t^{i}{}_{j}$,\\
$J_{a}^{i} = <v_{a},S t^{i}{}_{j}> x^{j}$. Using (\ref{newone}) we find
$$x^{l}(<v_{a},S t^{i}{}_{l}> L_{i}{}^{k}(x^{j})
- <O_{a}{}^{b},S t^{j}{}_{l}><v_{b},S t^{k}{}_{n}> x^{n}) = 0,$$
so that $L_{i}{}^{k}(x^{j})$
should be homogeneous of first order in $x$, which
suggests
$$L_{i}{}^{k}(x^{j}) =
<L_{i}{}^{k},S t^{j}{}_{l}> x^{l},\z L_{i}{}^{k} \in \U.$$

\subsubsection{Covariance of: $v f = v_{(1)}(f) v_{(2)}$}

Here: $v \in \U,\,f \in \fum$ and $v(f) = f^{(1)}<v,f^{(2)'}>$.
\paragraph{Covariance of $v(f)$ alone:}
\begin{equation}
\begin{array}{rcl}
\DA(v)\left(\DA f\right)
        & = & f^{(1)}<v^{(1)},f^{(2)'}> \otimes v^{(2)'}f^{(3)'}\\
        & = & f^{(1)}<v,f^{(3)'}> \otimes f^{(2)'}S(f^{(4)'})f^{(5)'}\\
        & = & f^{(1)} \otimes v(f^{(2)'})\\
        & = & \DA( v(f) ),\z\Box
\end{array}
\end{equation}
where we have used identity (\ref{usef}).
\paragraph{Covariance of the complete commutation relation:}
\begin{equation}
\begin{array}{rcl}
\DA v \cdot  \DA f
        & = & f^{(1)}<v^{(1)}{}_{(1)},f^{(2)'}>v^{(1)}{}_{(2)}
\otimes v^{(2)'}f^{(3)'}\\
        & = & f^{(1)}\underline{v^{(1)}{}_{(2)}} \otimes
        \underline{v^{(2)'}\,\widehat{v^{(1)}{}_{(1)}}(f^{(2)'})}\\
        & = & f^{(1)} v_{(2)}{}^{(1)}
\otimes v_{(1)}(f^{(2)'})\, v_{(2)}{}^{(2)'}\\
        & = & \DA(v_{(1)}(f) ) \DA(v_{(2)})\\
        & \stackrel{def}{=} & \DA(v f).\z\Box
\end{array}
\end{equation}
The underlined parts were rewritten using a compatability
relation between the right \A-coaction and the coproduct in \U:
\begin{equation}
v_{(2)}{}^{(1)} \otimes v_{(1)}(f^{(2)'})\, v_{(2)}{}^{(2)'} =
v^{(1)}{}_{(2)} \otimes v^{(2)'}\,\widehat{v^{(1)}{}_{(1)}}(f^{(2)'}),
\end{equation}
where ``$\widehat{\x}$''is the projector onto right-invariant vector
fields:\\ $\widehat{x} = S^{-1}(x^{(2)'}) x^{(1)} \in \smash$ for
$x \in \smash$.

\subsubsection{Covariance of:
$\partial _{i} f = \partial _{i}(f) + L_{i}{}^{j}(f) \partial _{j}$}

See section~\ref{covdfl}.
The main result was the following condition on $L_{i}{}^{j}$:
\begin{equation}
\left( L_{i}{}^{j}(f^{(2)'})
S^{2}t^{k}{}_{j} - S^{2}t^{l}{}_{i} \widehat{L_{l}{}^{k}}(f^{(2)'}) \right)
\otimes f^{(1)} = 0.
\end{equation}

\subsubsection{Covariance of: $J_{a}^{i} L_{i}{}^{k}(f) =
O_{a}{}^{b}(f) J_{b}^{k}$}

This proof is somewhat involved and we should keep in mind
that equation $v_{a} f = v_{a}(f) + O_{a}{}^{b}(f) v_{b}$ is already based on
$\DA$ being an algebra homomorphism; nevertheless, in several
steps:
\paragraph{$\DA$ is a homomorphism of}$\fum\cross\tqm$.
Proof on a function $f$:
\begin{equation}
\begin{array}{rcl}
\DA(J_{a}^{i} \partial _{i} f)
        & = & \DA(v_{a} f)\\
        & = & v_{a}^{(1)}f^{(1)} \otimes v_{a}^{(2)'}f^{(2)'}\\
        & = & <\underline{v_{a}^{(1)},
S t^{k}{}_{l}}> x^{l} \partial _{k} f^{(1)}
                \otimes \underline{v_{a}^{(2)'}} f^{(2)'}\\
        & = & <v_{a}, S t^{s}{}_{r}> x^{l}
\partial _{k} f^{(1)} \otimes S t^{r}{}_{l} S^{2}t^{k}{}_{s} f^{(2)'}\\
        & = & \underline{x^{l}} \partial _{k} f^{(1)}
\otimes \underline{S t^{r}{}_{l}
                <v_{a},S t^{s}{}_{r}>} S^{2}t^{k}{}_{s} f^{(2)'}\\
        & = & \DA J_{a}{}^{s} \,\DA(\partial _{s} f),\z\Box
\end{array}
\end{equation}
and also
\begin{equation}
\begin{array}{rcl}
\DA(J_{a}^{i} \partial _{i}( f) )
        & = & x^{j} \partial _{i}(f^{(1)})
\otimes S t^{s}{}_{j}<v_{a},S t^{r}{}_{s}> S^{2}t^{i}{}_{r} f^{(2)'}\\
        & = & \DA J_{a}{}^{r}\, \DA(\partial _{i})\left(\DA f\right)\\
        & = & \DA J_{a}^{r}\, \DA\left(\partial _{i}(f)\right).\z\Box
\end{array}
\end{equation}
\paragraph{A short aside,} checking
consistency of $O_{a}{}^{b}(f)J_{b}^{k}$ with
$\DA$ being an algebra homomorphism of $\fum$.
\begin{equation}
\begin{array}{rcl}
\underline{\DA(O_{a}{}^{b}(f) J_{b}^{i})}\, \DA(\partial _{i})
        & \stackrel{def}{=} & \DA(O_{a}{}^{b}(f) J_{b}^{i} \partial _{i})\\
        & = & \DA(O_{a}{}^{b}(f) v_{b})\\
        & = & \DA(O_{a}^{b}(f))\,\DA(v_{b})\\
        & = & \DA(O_{a}{}^{b}(f))\,\DA(J_{b}^{i} \partial _{i})\\
        & \stackrel{def}{=} &
        \underline{\DA(O_{a}{}^{b}(f))\,\DA(J_{b}^{i})}\,
\DA(\partial _{i}).\z\Box
\end{array}
\end{equation}
\paragraph{Synthesis:} Comparing
$$v_{a} f = v_{a}(f) + O_{a}{}^{b}(f) v_{b}$$ and $$J_{a}^{i} \partial _{i} f
= J_{a}^{i} \partial _{i}(f)
+ J_{a}^{i} L_{i}{}^{j}(f) \partial _{j}$$ we finally find:
\begin{equation}
\begin{array}{rcl}
\DA(J_{a}^{i} L_{i}{}^{k}(f) )
        & = & \DA(J_{a}^{i})\,\DA(L_{i}{}^{k}(f) )\\
        & = & \DA(O_{a}{}^{b}(f) J_{b}{}^{k} )\\
        & = & \DA(O_{a}{}^{b}(f) )\,\DA(J_{b}{}^{k}).\z\Box
\end{array}
\end{equation}
{\em Remark:} Given a linear operator $L_{i}{}^{j}:\fum \to  \fum$, satisfying
the appropriate consistency conditions,  equation
\begin{equation}
J_{a}^{i} L_{i}{}^{k}(f) = O_{a}{}^{b}(f) J_{b}{}^{k}
\end{equation}
could very well be used to give explicit covariant $x-x$ commutation
relations.

\subsection{Cartan Calculus on Quantum Planes}

So far we have only dealt with functions and (partial) derivatives
that we combined into an algebra of differential operators on the
quantum plane via commutation relations
\begin{equation}
\partial _{i} f = \partial _{i}(f) + L_{i}{}^{j}(f)
\partial _{j},\z \partial _{i}\in\tqm, \, f\in\fum.
\end{equation}
Now we would like to construct  differential forms through an
exterior derivative $\dl:\fum \to  \Lambda ^{1}(\fum)$ that is nilpotent
and satisfies the usual graded Leibniz rule.
Lie derivatives are introduced next, recalling that
they {\em act} on functions like the ordinary derivatives, that they
correspond to $\Li_{\partial _{i}}(f)
= \partial _{i}(f)$, and requiring that they
commute with the exterior derivative
$\Li_{\partial _{i}} \circ \dl = \dl \circ
\Li_{\partial _{i}}$. Just like it was the case for quantum Lie algebras, the
linear operator $L_{i}{}^{j}$  should also act like a Lie derivative, $i.e.$
we extend its definition from functions to forms by requiring that
it commute with $\dl$. Inner derivations
$\I_{\partial _{i}}$ are defined as graded linear
operators of degree -1, orthogonal to the natural
basis $\xi ^{i} := \dl(x^{i})$
of 1-forms: $\I_{\partial _{i}}(\xi ^{j})
= \delta _{i}^{j}$ --- in consistency with the Cartan identity
\begin{equation}
\Li_{\partial _{i}} = \I_{\partial _{i}} \dl + \dl \I_{\partial _{i}}
\end{equation}
that we want to postulate. For the exterior derivative of a function
we can choose between two expansions in terms of 1-forms
\begin{equation}
\dl(f) = \xi ^{i} a_{i} = b_{i} \xi ^{i}
\end{equation}
that we contract with $\I_{\partial _{j}}$ to find
\begin{equation}
\partial _{j}(f) = a_{j} = \I_{\partial _{j}}(b_{i} \xi ^{i})
\end{equation}
and
\begin{equation}
\dl(f) = \xi ^{i} \partial _{i}(f).
\end{equation}
The second expression has to wait while we
quickly  derive $x-\xi $-commutation
relations with the help of the first expression and the Leibniz
rule for $\dl$:
\begin{equation}
\begin{array}{rcl}
\dl f   & = & \xi ^{i}\partial _{i} f\\
        & = & \xi ^{i}\partial _{i}( f)
+ \xi ^{i} L_{i}{}^{j}(f) \partial _{j} = \\
\dl(f) + f \dl & = & \xi ^{i}\partial _{i}(f) + f \xi ^{j} \partial _{j},
\end{array}
\end{equation}
valid on any function and hence
\begin{equation}
f \xi ^{j} = \xi ^{i} L_{i}{}^{j}(f),\label{fxico}
\end{equation}
so that the second expression takes the (not so pretty) form
\begin{equation}
\dl(f) = \left(S L_{i}{}^{j} \circ \partial _{j}\right)(f),
\end{equation}
which, unlike in the quantum group case, does not simplify any further.
Lie derivatives and inner derivations along arbitrary first order
differential operators
$f^{i} \partial _{i},\x f^{i} \in \fum$, are introduced by the
following set of consistent equations:
\begin{eqnarray}
\I_{f^{i} \partial _{i}} & =
& f^{i} \I_{\partial _{i}},\\
\Li_{f^{i} \partial _{i}} & =
& \dl  \I_{f^{i} \partial _{i}} + \I_{f^{i} \partial _{i}} \dl,\\
\Li_{f^{i} \partial _{i}} & =
& f^{i} \Li_{\partial _{i}} + \dl(f^{i}) \I_{\partial _{i}},\\
\Li_{f^{i} \partial _{i}} \dl & = & \dl \Li_{f^{i} \partial _{i}}.
\end{eqnarray}
We will not give a complete set of commutation relations here because
the reader can easily obtain most of them from the quantum group treatment
simply by replacing $\Lio{i}{j} \to  L_{i}{}^{j}$. The problem of defining a
Lie bracket of vector fields on the quantum {\em plane} has,
however, not found a satisfactory solution yet.

\subsection{Induced Cartan Calculus}

We would like to complete the program started in section~\ref{indcal},
where we induced a calculus on the plane from the calculus on the
symmetry quantum group of that plane using a realization
$v_{a} \doteq J_{a}^{i} \partial _{i}$ of the
bicovariant group generators in terms of functions and derivatives on the
plane. From this expression we get the following two relations for
the Cartan generators on the   plane:
\begin{eqnarray}
\I_{v_{a}}  & \doteq & \I_{J_{a}^{i} \partial _{i}}
=  J_{a}^{i} \I_{\partial _{i}}\\
\Li_{v_{a}} & \doteq & \Li_{J_{a}^{i} \partial _{i}}
= J_{a}^{i} \Li_{\partial _{i}} + \dl(J_{a}^{i}) \I_{\partial _{i}}.
\end{eqnarray}
Commutation relations for the inner derivation with functions are
easily derived;
\begin{equation}
\I_{v_{a}} f = \Lio{a}{b}(f) \I_{v_{b}}
\end{equation}
and hence
\begin{equation}
J_{a}^{i} \I_{\partial _{i}} f = \Lio{a}{b}(f) J_{b}^{k} \I_{\partial _{k}}
\end{equation}
or, if a $\tilde{J}^{a}_{i}$ exists,
\begin{equation}
\I_{\partial _{i}} f = \tilde{J}^{a}_{i}
\Lio{a}{b}(f) J_{b}^{k} \I_{\partial _{k}},
\end{equation}
and finally
\begin{equation}
\I_{\partial _{i}} f = L_{i}{}^{k}(f) \I_{\partial _{k}}.
\end{equation}
Commutation relations for the Lie derivatives with functions can
now be calculated using the Cartan identity. We will present the
result of such a
computation for Wess-Zumino type linear planes (where $\tilde{J}^{a}_{i}$
exists):
\begin{equation}
\begin{array}{rcl}
\Li_{\partial _{i}} x^{l} & =
& \delta _{i}^{l} +
\underbrace{\tilde{J}^{a}_{i}
O_{a}{}^{b}(x^{l}) J_{b}^{k}}_{L_{i}{}^{k}(x^{l})} \Li_{\partial _{k}}\\
         &   & +\left( \dl(\tilde{J}^{a}_{i} O_{a}{}^{b}(x^{l}) J_{b}^{k}) -
                \tilde{J}^{a}_{i} \dl(O_{a}{}^{b}(x^{l})) J_{b}^{k} \right)
                \I_{\partial _{k}}.\label{innozero}
\end{array}
\end{equation}
The point is that not both
$O_a{}^b$ {\em and} $L_i{}^j$ can braid like Lie derivatives, \ie\ commute
with $\dl$.
Classically: $O_{a}{}^{b}(x^{l}) \to
\delta _{a}^{b} x^{l}$ and functions commute with functions and
forms so that the last term in the above equation vanishes.
The quantum case has a little surprise for us:
As was first discovered by B. Zumino \cite{pZ} through purely
algebraic considerations in the case of the
GL${}_{q}(2)$-plane, an inner derivation term is necessary in the
$\Li_{\partial }-x$ commutation relations in order to get consistency with
the undeformed Cartan identity.
Let us illustrate this at our standard example of the quantum plane
covariant under GL${}_{\frac{1}{q}}(n)$.

\subsubsection{Cartan Calculus for the 2-Dimensional Quantum Plane}

Using $x-\dl(x)$ commutation relations from (\ref{fxico})
\begin{eqnarray}
x \dl(x) & = & q^{2} \dl(x) x,\\
x \dl(y) & = & q \lambda \dl(x) y + q \dl(y) x,\\
y \dl(x) & = & q \dl(x) y,\\
y \dl(y) & = & q^{2} \dl(y) y,
\end{eqnarray}
we obtain
\begin{eqnarray}
\Li_{\partial _{x}} x & =
& 1 + q^{2} x \Li_{\partial _{x}} + q \lambda  y \Li_{\partial _{y}}
                + q \lambda
\dl(x) \I_{\partial _{x}} + \lambda ^{2} \dl(y) \I_{\partial _{y}},\\
\Li_{\partial _{x}} y & =
& q y \Li_{\partial _{x}} + \lambda  \dl(y) \I_{\partial _{x}}, \\
\Li_{\partial _{y}} x & =
& q x \Li_{\partial _{y}} + \lambda  \dl(x) \I_{\partial _{y}} ,\\
\Li_{\partial _{y}} y & =
& 1 + q^{2} y \I_{\partial _{y}} + q \lambda  \dl(y) \I_{\partial _{y}},
\end{eqnarray}
directly from (\ref{innozero}) after a lengthy computation.
Alternatively, we could have started with $\I_{\partial }-x$ commutation
relations
\begin{eqnarray}
\I_{\partial _{x}} x & =
& q^{2} x \I_{\partial _{x}} + q \lambda y \I_{\partial _{y}}, \\
\I_{\partial _{x}} y & = & q y \I_{\partial _{x}},\\
\I_{\partial _{y}} x & = & q x \I_{\partial _{y}} ,\\
\I_{\partial _{y}} y & = & q^{2} y \I_{\partial _{y}},
\end{eqnarray}
which have the great advantage that they have the exact same form
as the well-known $\partial -x$ relations. This also means that all of
our covariance considerations are still valid here.

\subsubsection{Extended Algebra for the $n$-Dimensional Quantum Plane}
\label{S:ndcc}

{\em Remark:} The explicit formulae in this section were obtained by B. Zumino
\cite{pZ}
using $r$-matrix techniques. They can alternatively be calculated with
the methods developed in this article.\\
Recall, that $J_a^i = v_a(x^i) = <v_a,St^i{}_j> x^j$.
For the covariance quantum group GL${}_{\frac{1}{q}}(n)$ we choose generators
\begin{equation}
v_{(nm)} = X^n{}_m \equiv <\frac{1}{\lambda}(1
\otimes 1 - \R^{21} \R^{12}) , t^n{}_m \otimes i\!d>,
\end{equation}
such that
\begin{equation}
J_{12} = <X_1,St_2> x_2 = -\eta_{12} x_2 = D_2 P_{12} x_2
\end{equation}
from (\ref{etagln}), in tensor product notation.
In particular we find:
\begin{equation}
\I_{X^n{}_m} = D^i{}_m x^n \partial_i,
\end{equation}
with $D =$ diag$\{q^{2n - 1}, \ldots , q^3, q\}$, and:
\begin{equation}
\Li_{X^n{}_m} = D^i{}_m(x^n \Li_{\partial_i} + \dl(x^n) \I_{\partial_i}).
\end{equation}
Here is a complete list of commutation relations for the Cartan Calculus:
\begin{eqnarray}
\I_{\partial_p} x^q & =
& q (\hat{r}^{-1})^{jq}{}_{ip} x^i \I_{\partial_j},\\
\I_{\partial_p} \dl(x^q) & =
& \delta^q_p - \frac{1}{q}
(\hat{r}^{-1})^{jq}{}_{ip} \dl(x^i) \I_{\partial_j},\\
\Li_{\partial_p} x^q & =
& \delta^q_p + q (\hat{r}^{-1})^{jq}{}_{ip} x^i \Li_{\partial_j}
     + \lambda (\hat{r}^{-1})^{jq}{}_{ip} \dl(x^i) \I_{\partial_j},\\
\Li_{\partial_p} \dl(x^q) & =
& \frac{1}{q} (\hat{r}^{-1})^{jq}{}_{ip} \dl(x^i) \Li_{\partial_j}.
\end{eqnarray}

\subsection*{Acknowledgements}

I would like to thank Paolo Aschieri, Chryssomalis Chryssomalakos,
Paul Watts and Bruno Zumino for many helpful conversations.
The explict formulae for the $n$-dimensional plane
given in the last section are due to B. Zumino.
Especially the material
presented in section~\ref{S:Cal} benefited considerably from
many discussions with Paul Watts.

This work was supported in part by the Director, Office of Energy
Research, Office of High Energy and Nuclear Physics, Division of High
Energy Physics of the U.S. Department of Energy under Contract
DE-AC03-76SF00098 and in part by the National Science Foundation under
grants PHY-90-21139 and PHY-89-04035.

\section*{Appendix}

Here we will give fairly detailed proofs of propositions $i)$ and $ii)$
and symbolic proofs of the related propositions $iii)$ through $v)$.
\paragraph{Proof of $i)$: }
We start by proofing a lemma about the relation of coactions in
\U\ and \A:
\begin{equation}
\begin{array}{rcl}
S^{-1}(x^{(2)'})<x^{(1)},S a> & =
& S^{-1}(x^{(2)'}) S(a_{(2)}) <x^{(1)},S a_{(1)}>
 a_{(3)}\\
        & = & S^{-1}(x^{(2)'}) x^{(1)}(S a_{(1)}) a_{(2)} \\
        & = & \widehat{x}(S a_{(1)}) a_{(2)}\\
        & = & <x , S a_{(2)}> S a_{(1)} a_{(3)}.\z\Box\label{alid}
\end{array}
\end{equation}
Another useful identity:
\begin{equation}
\fbox{$<x^{(1)},f> x^{(2)'}
= <x , f_{(2)}> f_{(1)} S(f_{(3)})$},\z\forall x \in \U,\x
f \in \A.\label{usef}
\end{equation}
\subparagraph{$i)$ ``$\Rightarrow$'':} Assume
$\DA \tq \subset \tq \otimes \A $, then for $\forall x \in \tq,\x S(a)
\in R$
\begin{equation}
0    =  <x^{(1)},S a> S^{-1}x^{(2)'}
         =  <x, S a_{(2)}> S(a_{(1)}) a_{(3)},
\end{equation}
so that $S a_{(2)} \otimes S(a_{(1)}) a_{(3)} \subset (R \oplus 1) \otimes \A$,
but $\epsilon (S a_{(2)}) S(a_{(1)}) a_{(3)} = \epsilon (S a) = 0$ and hence
$S a_{(2)} \otimes S(a_{(1)}) a_{(3)} \subset R \otimes \A$, or
\begin{equation}
\Delta ^{Ad}(a) \equiv a_{(2)} \otimes S(a_{(1)}) a_{(3)} \subset \widetilde{R}
\otimes \A.\z\Box
\end{equation}
\subparagraph{$i)$ ``$\Leftarrow$'':} Assume
$\Delta ^{Ad} \widetilde{R} \subset \widetilde{R} \otimes \A$, then again for
$\forall x \in \tq,\x a \in \widetilde{R}$
\begin{equation}
0 = < x, S a_{(2)}> S(a_{(1)}) a_{(3)} = <x^{(1)},S a> S^{-1}x^{(2)'},
\end{equation}
so that $x^{(1)}
\otimes S^{-1} x^{(2)'}\subset (\tq \oplus 1) \otimes \A$; with
$0 = <x,1> = <x^{(1)},1> x^{(2)'}$  from (\ref{usef}) that gives
$x^{(1)} \otimes S^{-1}x^{(2)'} \subset \tq \otimes \A$ and also
\begin{equation}
\DA x = x^{(1)} \otimes x^{(2)'} \subset \tq \otimes \A.\z\Box
\end{equation}

\paragraph{Proof of $ii)$: }
\subparagraph{$ii)$ ``$\Rightarrow$'': } For all $x \in \tq$, $a \in \A$ and
$r \in R$ assume $\Delta  x \in \U \otimes (\tq \oplus 1)$, then
\begin{equation}
<x, a r> = <\Delta  x ,a \otimes r > = 0,
\end{equation}
which implies $a r \in (R \oplus 1)$ or, taking into account that
$\epsilon (a r) = \epsilon (a) \epsilon (r) = 0$,
\begin{equation}
a r \in R.\z\Box
\end{equation}
\subparagraph{$ii)$ ``$\Leftarrow$'': } Assume that for all $x \in \tq,\x r
\in R$ there exists a $r' \in R$ such that $r' = a r$; then we find
\begin{equation}
0 = <x,r'> = <x,a r> = <\Delta  x, a \otimes r>
\end{equation}
which can be restated as
\begin{equation}
\Delta  x \in \U \otimes (\tq \oplus 1).\z\Box
\end{equation}
\paragraph{Symbolic proof of $iii)$: }
\begin{equation}
0 = <\tq^{\perp } \otimes i\!d
,(S \otimes i\!d)\circ \Delta ^{Ad} \widetilde{R^{\perp }}> =
<(i\!d \otimes S^{-1}) \circ
\DA \tq^{\perp }, S \widetilde{R^{\perp }} \otimes i\!d>
\end{equation}
\paragraph{Symbolic proof of $iv)$: }
\begin{eqnarray}
\lefteqn{0 = <R^{\perp },\tq^{\perp }> =  <R^{\perp }, \U \tq^{\perp }>
= }\nonumber\\
& & <\Delta  R^{\perp },\U
\otimes \tq^{\perp }> =
<\A \otimes (R^{\perp } \oplus 1), \U \otimes \tq^{\perp }>
\end{eqnarray}
\paragraph{Symbolic proof of $v)$: }
\begin{eqnarray}
\lefteqn{0 = <R^{\perp },\tq^{\perp }> = <R^{\perp }, \tq^{\perp } \U> =
}\nonumber\\
& & <\Delta  R^{\perp },\tq^{\perp }
\otimes \U > =
<(R^{\perp } \oplus 1) \otimes \A,\tq^{\perp } \otimes \U>
\end{eqnarray}

\end{document}